\documentclass[12pt]{article}
\usepackage{amsmath}
\usepackage{graphicx}
\usepackage{enumerate}
\usepackage[numbers]{natbib}	
\usepackage{url} % not crucial - just used below for the URL 

\newcommand{\blind}{1}

\usepackage{booktabs}
\usepackage{adjustbox}
\usepackage{bookmark}
\usepackage{bm}
\newenvironment{proof}{\paragraph{Proof:}}{\hfill$\square$}
\usepackage{subcaption}
\usepackage{float}
\usepackage{enumerate}% http://ctan.org/pkg/enumerate
\usepackage{soul}
\usepackage{amsmath, amssymb} 
\usepackage{mathrsfs}
\usepackage{subcaption}
\usepackage{xcolor}

\usepackage[framemethod=tikz]{mdframed}

\usepackage{mhequ}
\usepackage{xcolor}
\usepackage{hyperref}

\newtheorem{theorem}{Theorem}

\newtheorem{remark}{Remark}

\makeatletter

\DeclareMathOperator{\No}{\text{N}}
\newcommand{\prox}{\t{prox}_{\lambda g} }
\newcounter{example}[section]
\newenvironment{example}[1][]{\refstepcounter{example}\par\medskip
   \noindent \textbf{Example~\theexample. #1} \rmfamily}{\medskip}

\DeclareMathOperator*{\argmin}{arg\,min}
\newcommand{\pr}{\text{pr}}
\newcommand{\Beta}{\boldsymbol{\beta}}

\newcommand{\T}{\textrm{T}}

\renewcommand{\t}[1]{\text{#1}}

\usepackage{xcolor}

\newcommand{\be}{\begin{equation*}
        \begin{aligned}}

\newcommand{\ee}{ \end{aligned}
        \end{equation*}
}

\newcommand{\bel}{\begin{equation}
        \begin{aligned}}

\newcommand{\eel}{ \end{aligned}
        \end{equation}
}

\usepackage{float}

\begin{document}

\def\spacingset#1{\renewcommand{\baselinestretch}%
{#1}\small\normalsize} \spacingset{1}

%%%%%%%%%%%%%%%%%%%%%%%%%%%%%%%%%%%%%%%%%%%%%%%%%%%%%%%%%%%%%%%%%%%%%%%%%%%%%%

\if1\blind
{
        \title{Bayesian Inference Using the Proximal Mapping:\\ 
        Uncertainty Quantification Under \\ Varying Dimensionality}
\author{Maoran Xu\footnote{Department of Statistics, University of Florida, {maoranxu@ufl.edu}}
\quad Hua Zhou\footnote{Departments of Biostatistics and Computational Medicine, University of California, { huazhou@ucla.edu}}
\quad Yujie Hu \footnote{Department of Geography, University of Florida, {yujiehu@ufl.edu}}
\quad Leo L. Duan\footnote{Department of Statistics, University of Florida, {li.duan@ufl.edu}}}
  \maketitle
} \fi

\if0\blind
{
  \bigskip
  \bigskip
  \bigskip
  \begin{center}
    {\LARGE\bf Bayesian Inference Using the Proximal Mapping: 
        Uncertainty Quantification Under  Varying Dimensionality}
\end{center}
  \medskip
} \fi

\bigskip
\begin{abstract}
 In statistical applications, it is common to encounter parameters supported on a varying or unknown dimensional space. Examples include the fused lasso regression, the matrix recovery under an unknown low rank, etc. Despite the ease of obtaining a point estimate via optimization, it is much more  challenging to quantify their uncertainty. In the Bayesian framework, a major difficulty is that if assigning the prior associated with a $p$-dimensional measure, then there is zero posterior probability on any lower-dimensional subset with dimension $d<p$. To avoid this caveat, one needs to  choose another dimension-selection prior on $d$, which often involves a highly combinatorial problem. To significantly reduce the modeling burden, we propose a new generative process for the prior: starting from a  continuous random variable such as multivariate Gaussian,
                we transform it into a varying-dimensional space using the proximal mapping.
                This leads to a large class of new Bayesian models that can directly exploit the popular frequentist regularizations and their algorithms, such as the nuclear norm penalty and the alternating direction method of multipliers, while providing a principled and probabilistic uncertainty estimation.
                We show that this framework is well justified in the geometric measure theory,
                and enjoys a convenient posterior computation via the standard Hamiltonian Monte Carlo.  We demonstrate its use in the analysis of the dynamic flow network data.
\end{abstract}

\noindent%
\vfill

\newpage
\spacingset{1.2} % DON'T change the spacing! 
%% Here are the title, author names and addresses
      
\section{Introduction}

Modern statistical applications often involve data that are high dimensional. To allow signal recovery under a relatively low sample size, one often needs to assume that the parameter $\theta\in \mathbb{R}^p$ in fact lies in/near some lower dimensional space. Commonly used assumptions include sparsity \citep{meinshausen2006high,zhao2006grouped,meier2008group,bickel2009simultaneous}, low rank \citep{shen2008sparse,ji2010robust}, geometric constraints \citep{saarela2011method,goodall1999projective}, etc. In most cases, the dimensionality $d$ is unknown. For example, we usually do not know the exact rank in the low-rank matrix factorization.
        
Bayesian framework provides a principled way to quantify the uncertainty on those models. A potential caveat is that if the assigned prior  is associated with a $p$-dimensional continuous measure, then there is zero posterior probability allocated on any of the lower-dimensional subsets with dimension $d<p$. Instead, from a generative perspective, one should  first choose a discrete prior to select $d$, then generate $\theta$ within the chosen space. For example, the spike-and-slab prior \citep{mitchell1988bayesian} assigns a binomial distribution on $d$ as the number of non-zero coefficients in the variable selection problem; the Bayesian adaptive regression spline uses a Poisson prior on the number of knots $d$, which determines the rank of the spline matrix \citep{dimatteo2001bayesian}. On the other hand, the discrete prior creates a highly combinatorial problem, and existing estimation methods such as the Reversible-jump Markov chain Monte Carlo \citep{green2009reversible} are not very efficient to explore the high posterior probability region.

An appealing alternative is to avoid specifying any low-dimensional prior, but to induce a prior for $\theta$ with the measure in $\mathbb{R}^p$ yet having the mass concentrated near some low-dimensional sets. Specifically, the key is to re-parameterize the parameter $\theta$ as some transformation of a sparse vector $\beta$, and then assign simple continuous shrinkage prior on $\beta$ to favor near-zero values. For example, in spline regression, one uses $\beta$ as the sparse
weights in the linear combination of some basis functions. In this category, there is a rich literature  covering tasks of variable selection \citep{park2008bayesian,carvalho2009handling,rovckova2018spike}, matrix decomposition \citep{bhattacharya2011sparse,legramanti2020bayesian}, functional data analysis \citep{shin2020functional}, covariance estimation \citep{li2019graphical,kastner2019sparse}, among others. 

Clearly, this strategy has its limitations ---  when we cannot re-parameterize the low-dimensional sets of $\theta$, the prior specification becomes awkward. This is not uncommon. For example, the fused lasso \citep{tibshirani2005sparsity} is a frequentist regularization very popular in the image/signal processing, which assumes sparsity not only in the parameter $\theta\in \mathbb{R}^p$, but also in the $(p-1)$ increments between the neighboring elements $(\theta_{j+1}-\theta_j)$. Although we could imagine assigning some shrinkage prior on $\beta=D\theta\in \mathbb{R}^{2p-1}$ with $D$ the  corresponding matrix, such a prior is ill-defined: as $D$ is not invertible,  we cannot compute $\theta$ from $\beta$; as $\beta$ resides in the column span of $D$, it has a dimension $p$, which is less than $(2p-1)$ --- therefore, the shrinkage prior one blindly assigned would be in fact an incomplete density for a degenerate measure, making it difficult to calibrate the hyper-parameters within and assess the effects of the prior regularization.

Motivated to generalize the Bayesian approaches for handling most of the low-dimensional regularizations (including potentially complicated ones),
while avoiding the caveats of having to explicitly specify a discrete prior, we consider a ``projection''-style approach. Starting from a continuous prior for $\beta$ with measure in $\mathbb{R}^p$, we transform it into $\theta$ using a special mapping, so that $\theta$ has an induced prior on several low-dimensional sets. The projection idea was previously considered in several cases, such as the mixture of components with different dimensions  \citep{petris2003geometric}, the isotonic regression \citep{dunson2003bayesian}, monotone curve fitting  \citep{lin2014bayesian} and more generally, constrained space modeling \citep{patra2018constrained}. Nevertheless, in this article, we explore a much more general transformation known as the proximal mapping --- it not only includes common Euclidean projection to a constrained set, but also useful non-projection transformation such as soft-thresholding, nuclear norm control, set expansion, etc. This mapping has been well studied in the optimization literature, with appealing properties that are  convenient for canonical Bayesian inference, such as in the concentration of measure and convenient computation via the Hamiltonian Monte Carlo. We will carefully justify this prior via the geometric measure theory and demonstrate the strengths via several examples.

\section{Method}
        \label{sec: method}
        \subsection{Background on the Proximal Mapping}
        We first provide a brief review on the proximal mapping and motivate its use as a transformation tool. Let $\theta$ be the parameter of interest in a certain space $\Theta$, with $\Theta \subseteq \mathbb{R}^p$. With another parameter $\beta\in {\boldsymbol \beta} \subseteq \mathbb{R}^p$, the \emph{proximal mapping} is a transform of $\beta$ to $\theta$:
                \bel\label{eq:proximal_mapping}
                \theta=\t{prox}_{\lambda g}(\beta):=\argmin_{z\in \Theta}\left \{ \lambda g(z)+ \frac{1}{2}\|z-\beta\|_2^2\right\},
        \eel
  where $g$ is a lower semi-continuous and convex function, and $\lambda>0$ is a scalar as a hyper-parameter. This effectively induces a parameter space:
  \be
  \Theta_{\lambda g} =  \{ \t{prox}_{\lambda g}(\beta): \beta\in \mathbb{R}^p\}.
  \ee
  
  For an intuitive understanding, the proximal mapping could be viewed as a generalized projection. Given a constrained set $C$, we 
  can choose  $g(z)=  \mathcal X_{C}(z)$, the characteristic function of a constrained set taking value $0$ if $z\in C$, or $\infty$ if $z\not\in C$. The mapping becomes $\theta=P_C(\beta)=\arg\min_{z\in C} \| z-\beta\|^2_2$, the Euclidean projection of $\beta$ into the set $C$. Furthermore, we can replace $\mathcal X_{C}$ with other function for $g$, leading to a wider class of transformation.
\setcounter{example}{0}
\begin{example}[Soft thresholding]
  Perhaps the most famous example is  $g(z)=\|z\|_1$  from lasso \citep{tibshirani1996regression}. It has a closed-form proximal mapping known as the soft-thresholding operator $\t{prox}_{\lambda g}(\beta)= \t{sign}(\beta) \max(|\beta|-\lambda,0)$, with all operations carried out element-wise. The induced parameter space $\Theta_{\lambda g}$ is in fact the union of multiple sets with varying dimensions: $\{\theta \in \mathbb{R}^p: \theta_j=0 \t{ for } j \in S, |S|=d\}$, where $S$ is some index set and $d\in (0,1,\ldots, p)$, each is a Euclidean subspace of dimension $(p-d)$ --- conveniently, we do not need to explicitly specify the dimension $d$, since it is automatically induced through the transformation.
\end{example}

This suggests that the proximal mapping can be used as a convenient tool to develop priors on lower-dimension subsets. We now list a few useful proximal mappings in Table~\ref{tab:tab_prox}. In addition, the proximal mapping allows us to easily consider multiple constraints or $g$ functions, since the intersection of convex sets and summation of  convex functions are still convex. The general form can be computed using the alternative direction of method of multipliers algorithm \cite{bertsekas2014constrained}, and we will demonstrate one case in the data application.
For example, consider $\theta$ being sparse while constrained in some convex set; this would be challenging to model for conventional approaches due to the lack of reparametrization.

\begin{table}[htbp]
\spacingset{1}
\caption{Some useful proximal mappings.  \label{tab:tab_prox}}
\centering
\small
\begin{tabular}{p{25mm} | p{40mm} | p{40mm}  | p{30mm} }
% \hline
Space of $\beta$ & $g(z)$ & $\t{prox}_{\lambda g}(\beta)$  & Usage \\
\hline
$\mathbb{R}^p$ & $\mathcal X_C$, $C$ convex set & $P_C(\beta)$ &  Projection to a set [See Table 6.1 of \cite{beck2017first} for an expanded list] \\
\hline
$\mathbb{R}^p$ & $\|z\|_1$ & $\t{sign}(\beta) \max(|\beta|-\lambda,0)$, computed element-wise & Sparsity \\
\hline
$\{\beta\in \mathbb{R}^{k\times k}$, \newline  positive semi-definite\}  & $\|Z\|_*$,  nuclear norm & 
 $U \Lambda_0 V^{\T}$, with $\beta=U \Lambda V^{\T}$ the singular value decomposition, $\left(\Lambda_{0}\right)_{i i}=\max (0, \Lambda_{i i}-\lambda)$ & Low rank
\\
\hline
$\mathbb{R}^{m\times n}$ & $\|Z\|_{2,1}=\sum_{i}\sqrt{\sum_{j} Z_{ij}^2}$ & $[\beta_i\max(1-\frac{\lambda}{\|\beta_i\|_2},0)]_{i=1}^m$ with $\beta_i$ as the $i$th row & Row / group sparsity
\\
\hline
$\mathbb{R}^p$ & $\|Dz\|_1$ with \newline $D\in\mathbb{R}^{k\times p}$  & Solvable via the alternating direction method of multipliers & Fused lasso, \newline convex clustering \\
\hline
$\mathbb{R}^p$ & $\t{dist}_C(z)=\inf_{x\in C}\|z-x\|_2$, \newline distance to a set & $aP_C(\beta) + (1-a)\beta$  with \newline $a=\min\{\lambda/\t{dist}_C(\beta),1\}$
& Set expansion to $C$ \\
\hline
\end{tabular}
\end{table}
\subsection{Proximal Prior}
We now use the above in a Bayesian modeling framework. Suppose we have  data generated from a likelihood $L(y;\theta)$, where we want to assign a prior on $\theta$ in some space with dimensionality smaller or equal to $p$. We use the following generative process for $\theta$:
\bel \label{eq:proximal_prior}
&\beta \sim \Pi^0_{\bm\beta},\\
& \lambda \sim \Pi^0_{\bm\lambda},\\
& \theta = \t{prox}_{\lambda g}(\beta),
\eel
where $\Pi^0_{\bm\beta}$ is a  continuous distribution in $\mathbb{R}^p$, such as the non-degenerate Gaussian $\beta\sim \No(\mu,\Sigma)$ and we use $ \Pi^0_{\bm\lambda}$ to denote a generative distribution for $\lambda>0$. 

Here $g$ is a convex and lower-semicontinuous function such as those in Table 1. Potentially, $g$ could be known up to some other hyper-parameter $\gamma$; in that case, we denote it by $g_\gamma$ and use $\Pi^0_{\bm \gamma}$ as the prior for $\gamma$. For a clear notation, we use bold subscript such as in $\Pi^0_{\bm\theta}$ as a book-keeping index to refer to the variable whose prior is being defined.

%  \maoran{In the following, we keep the notation $\Pi^0_{\bm x}(x)$ to denote the prior distribution of random variable $x$, with a bold symbol of this variable on the subscript. We use $\Pi(A\mid y)$ to denote the posterior probability of event $A$ given the data $y$}.
 
It is not hard to see that  $\lambda g(z)+{2}^{-1}\|z-\beta\|_2^2$, as the combination of the convex $g$ and a quadratic term,  is strictly convex with a unique minimizer. Therefore, each $\beta$ maps to a unique $\theta$, hence we have a measurable mapping, which means we have a valid prior distribution for $\theta$ using \eqref{eq:proximal_prior}. { We denote the conditional prior distribution for $\theta$ as $\Pi^0_{\bm\theta}(\theta\mid \lambda, \gamma)$, and its marginal distribution as $\Pi^0_{\bm\theta}(\theta)$ after integrating out $\gamma$ and $\lambda$. For convenience, we will refer to either form as a ``proximal prior''. 

%We now illustrate the varying dimensional properties induced by the proximal prior. 
%On the surface, it might be tempting to use the Bayes' theorem to compute the posterior as
%\(
%\Pi(\theta \mid y)   = z^{-1} \Pi^0_{\bm\theta}(\theta) L(y;\theta),
%\)
% where $L(y;\theta)$ denotes the likelihood and $z$ is a normalizing constant that only depends on $y$. However, under the condition that $\theta=\prox(\beta)$, there is an issue that needs to be more carefully addressed: neither $\Pi^0_{\bm\theta}$ nor $\Pi(\theta \mid y)$ is a simple density, since one cannot simply integrate them over $\mathbb{R}^p$, due to the potentially varying dimensionality in the output of $\prox$. We will rigorously address this later using the geometric measure theory; for now, we use simple statistical concepts (sets and conditional probabilities) to show some useful properties.

We first show that, a proximal prior can produce a convenient equivalence to a hierarchical prior of first selecting a low-dimensional set and then assigning a conditional density within this set. We denote the space induced by $\prox(\Beta)$ as $\Theta$,  and assume that it can be  partitioned into $\Theta= \Theta^{0}\cup \Theta^{1} \cup \ldots\cup \Theta^{p}$, where $\Theta^k$ denotes a $k$-dimensional subset of $\Theta$, and $\Theta^{j}\cap \Theta^{k} =\varnothing$ if $j\neq k$ (this can be achieved even if a higher dimensional set $ \tilde\Theta^k$ overlaps/contains a lower-dimensional set $\Theta^j$, we set $\Theta^k= \tilde\Theta^k \setminus \cup_{j=1}^{k-1}\Theta^j$). Then the prior kernel (a mix of density and mass functions) evaluated at $\theta=t$ can be written as:
\bel\label{eq:prior_kernel}
\Pi^0_{\bm\theta}(t) = \sum_{k=0}^{p}\Pi^0_{\bm\theta}(t\mid \theta\in \Theta^k){\bf 1}(t\in \Theta^k)\t{pr}(\theta\in \Theta^k) ,
\eel
where $\sum_{k=0}^p\t{pr}(\theta\in \Theta^k) =1$ and $\Pi^0_{\bm\theta}(t\mid \theta\in \Theta^k)$ is a conditional density that integrates to $1$ over $t\in \Theta^k$ using an appropriate $k$-dimensional integral with respect to some proper measure $\lambda^k$, denoted by $\int_{\Theta^k} \Pi^0_{\bm\theta}(t\mid \theta\in \Theta^k) \lambda^k(\textup{d}t)=1$. The integral and measure will be formally defined in the theory section. 

Therefore, from a generative view, the above can be understood as first picking a set $\Theta^k$ with probability $\t{pr}(\theta\in \Theta^{k})$, then drawing a value $t$ within the space of $\Theta^k$. This includes those corner cases where $\prox$ cannot map to some dimensional sets: that is, for some $k$'s, we can have $\t{pr}(\theta\in \Theta^k)=0$. 

Accordingly, with $L(y;\theta)$ the likelihood, the posterior of $\theta$ can be derived as:
\bel
\label{eq:posterior_kernel}
\Pi(\theta =t \mid y)   =   \sum_{k=0}^{p}
\underbrace{ z_k^{-1}L(y;t)  \Pi^0_{\bm\theta}(t\mid \theta\in \Theta^k) {\bf 1}(t\in \Theta^k)}_{\Pi(\theta =t\mid \theta\in \Theta^k,y )}
\underbrace{
 \frac{z_k\t{pr}(\theta\in \Theta^k)}{\sum_{k=0}^dz_k\t{pr}(\theta\in \Theta^k)}}_{\t{pr}(\theta\in \Theta^k \mid y ) },
\eel
where $z_k = \int_{\Theta^k} L(y;\theta=t)  \Pi^0_{\bm\theta}(t\mid \theta\in \Theta^k) \lambda^k(\textup{d}t)$. We assume posterior propriety almost everywhere, such that $z_k<\infty$ for all $k:\t{pr}(\theta\in \Theta^k)>0$.%; and for those  $k:\t{pr}(\theta\in \Theta^k)=0$ and $t\in \Theta^k$, we fix $z_k:=0$ and $ \Pi(\theta =t \mid y):=0$.

The proximal priors simplify these procedures.
Using the transformation $\theta=\prox(\beta)$, for any measurable set $\mathcal A\in \Theta$,
\be
& \t{pr}(\theta\in \mathcal A \mid y)  \\  
= & 
{
  \sum_{k=0}^{p}
{ z_k^{-1}\left\{ \int_{\mathcal A\cap \Theta^k} L(y;t)  \Pi^0_{\bm\theta}(t\mid \theta\in \Theta^k) \textup{d} t \right\}
 \frac{z_k\t{pr}(\theta\in \Theta^k)}{\sum_{k=0}^dz_k\t{pr}(\theta\in \Theta^k)}}}
\\
\stackrel{(a)}=&  \sum_{k=0}^{p} z_k^{-1}\left[ \int_{\prox^{-1}(\mathcal A\cap \Theta^k)}L\left \{ y; \prox(b)\right \} \frac{\Pi^0_{\bm\beta}(b) {{\bf 1}[\prox(\beta)\in \Theta^k]}}{\t{pr}[\prox(\beta)\in \Theta^k]} \textup{d} b \right ] \frac{z_k\t{pr}(\theta\in \Theta^k)}{\sum_{k=0}^dz_k\t{pr}(\theta\in \Theta^k)} \\
=&  \frac{1}{\sum_{k=0}^dz_k\t{pr}(\theta\in
\Theta^k)} \sum^{p}_{k=0}\int_{\prox^{-1}(\mathcal A\cap \Theta^k)} L\left \{ y;\prox(b)\right \} {\Pi^0_{\bm\beta}( b)\textup{d} b },
\ee
where in $(a)$ we mean that $\Pi^0_{\bm\theta}(t\mid \theta\in \Theta^k)$ contains the Jacobian term in the change-of-variables $t=\prox(b)$ (details of the Jacobian calculation provided in the theory section).
% is due to the integral can be viewed as the conditional expectation $\mathbb{E}_{\bm \theta}\{L(y;\theta) {\bf 1}(\theta\in \mathcal A) \mid \theta\in \Theta^k\} = \mathbb{E}_{\bm \beta}[ L\{y;\prox(\beta)\} {\bf 1}\{\prox(\beta)\in \mathcal A\} \mid
%\prox(\beta)\in \Theta^k]$, where we know from the generative model that the conditional density $\Pi^0_{\bm\beta}\{b\mid \prox(\beta)\in \Theta^k\} = {\Pi^0_{\bm\beta}(b)\bf{1}\{\prox(\beta)\in \Theta^k\} }/{\t{pr}\{\prox(\beta)\in \Theta^k\}}$. 
%When $\Pi^0_{\bm\theta}(t\mid\theta\in\Theta^k)$ is well-defined,
 At any given $\beta=b$, we can omit the integral and summation, and  obtain a remarkably simple posterior density of $\beta$:
\bel\label{eq:post_density_beta}
\Pi(\beta=b \mid y)\propto L\left \{ y;\prox(b)\right \} {\Pi^0_{\bm\beta}( b)}.
\eel 
\begin{remark}
To clarify, although the hierarchical form of $\Pi^0_{\bm \theta}$ provides a nice interpretation to our proximal prior, such an equivalence is not strictly necessary for the proximal modeling framework to work. To be rigorous, the above equivalence requires a few regularity conditions, to be formalized in the theory section.
\end{remark}

Therefore, compared to \eqref{eq:posterior_kernel},  the posterior density \eqref{eq:post_density_beta} is much easier for Bayesian applications. This also suggests a new strategy of ``data augmentation using optimization'' [instead of marginalization as in \cite{tanner1987calculation}] --- if we can write the parameter $\theta$ as some proximal mapping from $\beta$, then we can sample $\beta$ first as an augmented variable; after sampling, we compute $\theta=\prox(\beta)$ and discard the information from $\beta$.

%\maoran{
%We now illustrate with several examples that the proximal priors can be applied in classic varying dimensional problems such as sparse, low-rank modeling, and show that the proximal priors can achieve the conventional priors with straightforward modeling process.}

We now use one example to illustrate the equivalence.

\begin{figure}[H]
        \centering
        \begin{subfigure}[t]{.4\textwidth}
        \centering
\includegraphics[width=\textwidth]{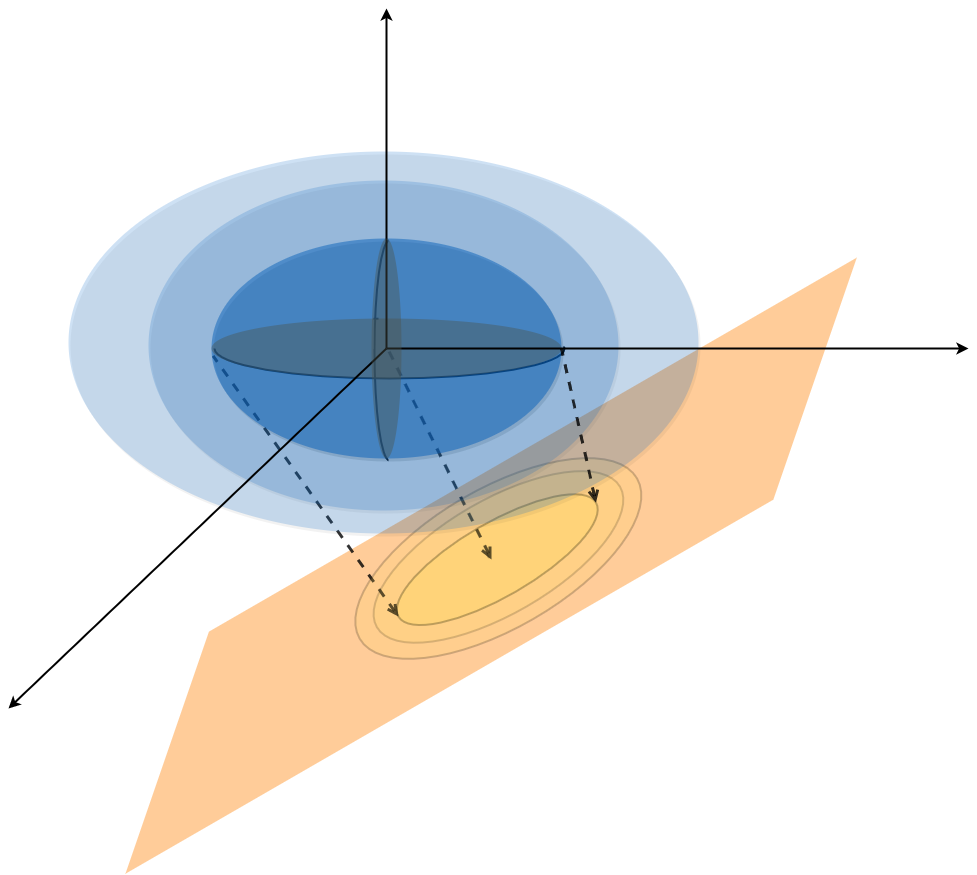}
\caption{Given $\t{rank}(A)=1$, the mapping $\theta=\prox(\beta)$ creates a two-dimensional prior on the hyperplane (orange).}
        \end{subfigure}
        \qquad
        \begin{subfigure}[t]{.4\textwidth}
        \centering
\includegraphics[width=\textwidth]{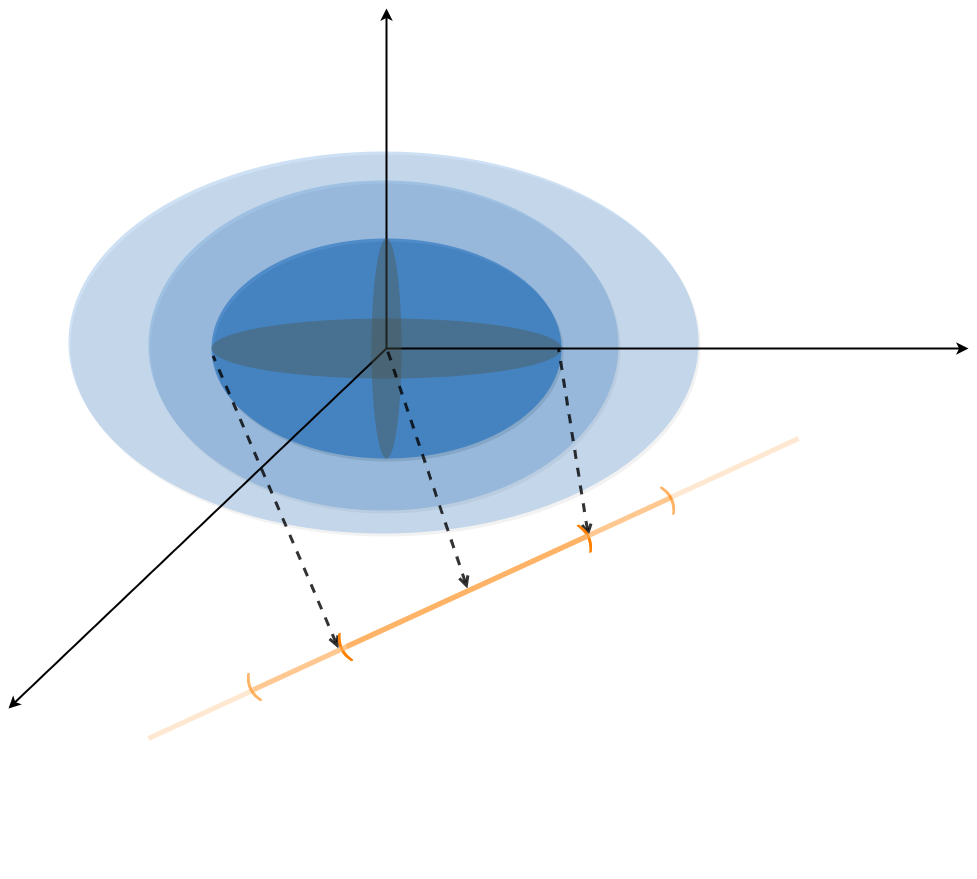}
\caption{Given $\t{rank}(A)=2$, the mapping $\theta=\prox(\beta)$ creates a one-dimensional prior on the line (orange).}
        \end{subfigure}
        \caption{\label{fig: proj_ellipsoid}
        Illustrative example of constructing a prior on an affinely constrained set $C=\{\theta:A^\textrm{T}\theta=b\}$. A  challenge arises when the rank$(A)$ is unknown, the dimensionality of the prior is unknown. The proximal prior bypasses this hurdle by transforming a continuous prior (blue) into the constrained space (orange), without the need to explicitly specify the dimensionality.}
\end{figure}

\begin{example}[Affinely constrained prior under varying rank] 
  Suppose we want to  assign a prior for $\theta$ in a set of  affine constraints $C=\{\theta\in \mathbb{R}^p: A^\T \theta =b\},$ where $A^\T\in \mathbb R^{m\times p}$ is another  parameter, with $m<p$ and $b\in \t{Col}(A^{\T})$ the column space of $A^{\T}$ (so that $C$ is not empty). 
  Since $A$ is not fixed, we do not know the rank of $A$, hence not the dimensionality of $C$.  Using the proximal prior with $g_A(z) = 
  0$ if $A^{\T}\theta=b$, 
  $g_A(z) =\infty$ otherwise (hence $\t{prox}_{\lambda g_A}$ is invariant to any finite value of $\lambda>0$), and $\beta \sim  \No(\mu,\Sigma)$, we have a closed-form proximal mapping \[\theta=\t{prox}_{\lambda g}(\beta)= \beta-A(A^\T A)^{-}(A^\T\beta-b),\] where $(\cdot)^-$ is the Moore-Penrose inverse.  We illustration  this mapping in Figure~\ref{fig: proj_ellipsoid}.

 The $\theta$-marginal proximal prior is 
%the a continuous mixture  over different value of $A$, or equivalently, 
a discrete mixture over different rank of $A$:
\begin{align*}
\Pi_{ \bm \theta}^{0}(\theta) & =\int \Pi_{\bm\theta}^{0}\left(\theta \mid A\right) \Pi^0_{\bm A}(A) \textup{d}A \\
 & =   \sum_{d=0}^p \text{pr}^0(\text{rank}(A)=d) \underbrace{\frac{\int_{A:\text{rank}(A)=d} \Pi_{\bm\theta}^{0}\left(\theta \mid A\right) \Pi^0_{\bm A}(A) \textup{d}A }{ \int_{A:\text{rank}(A)=d} \Pi^0_{\bm A}(A) \textup{d}A }}_{\Pi_{\bm\theta}^{0}\left[\theta \mid \t{rank}(A)=d\right]}
\end{align*}
 and $\Pi_{\bm\theta}^{0}\left(\theta \mid A\right)$ is the degenerate Gaussian density with mean $A\left(A^{\mathrm{T}} A\right)^{-1} b + P_{A^{\perp}} \mu$ and covariance $P_{A^{\perp}} \Sigma P_{A^{\perp}}$, where $P_{A^{\perp}} = I - A\left(A^{\mathrm{T}} A\right)^{-1}A^{\rm T}$.  Although  the summation may not have a closed-form, the weights and conditional density can be tractable in applications.

For illustration, we consider a Bayesian envelope  linear regression for multivariate response   $Y_i\in \mathbb{R}^{p}$:
\begin{align*}
  &  Y_i = \mu+ \tilde \Theta X_i + \epsilon_i, \quad \epsilon_i \stackrel{\text{iid}}\sim \text{N}  \big ( 0_p, \; \Gamma\Omega \Gamma^{\rm T} +  \Gamma_*\Omega_* \Gamma_*^{\rm T} \big )
 \\
& \theta = \Gamma \eta, 
\end{align*} 
for $i=1,\ldots,n$, each covariate $X_i\in\mathbb{R}^m$ (with $p\le m$), and the noise $\epsilon_i\in \mathbb{R}^{p}$; $[\Gamma \; \Gamma_*]$ together from a $p\times p$ orthonormal matrix, with $\Gamma$ a $p \times u$ sub-matrix, $\eta\in\mathbb{R}^{u\times m}$ full rank, and both $\Omega$ and $\Omega_*$ positive definite matrices. The regression coefficient matrix $\tilde \Theta\in\mathbb{R}^{p\times m}$ is of rank $u$, with $u$ unknown. The motivation is that by making $\tilde \Theta X_i$ in the  subspace spanned by the leading eigenvectors of the covariance, $\Gamma_*^{\rm T} Y_i$ is small in magnitude and independent from $X_i$, leading to a sufficient dimension reduction \citep{cook2010envelope}. For Bayesian inference, \cite{khare2017bayesian} proposed to use a matrix-Bingham prior on $\Gamma$ with a pre-specified $u$, so that it has conjugate forms in a Gibbs sampler for posterior computation. On the other hand, since $\Gamma$ is in an orthogonal and low-rank space, it is difficult to generalize to other forms of prior such as letting $u$ vary.

Using the affine constraint proximal mapping, we can bypass these challenges.
 We reparameterize
  $\Gamma_*\Omega_* \Gamma_*^{\rm T} = AA^{\rm T}$,
  $\Gamma\Omega \Gamma^{\rm T} = P_{A^{\perp}} W P_{A^{\perp}}^{\rm T}$
   with rank$(A)=p-u$, $W$ positive definite, and $\tilde\Theta$ by a linear constraint $A^{\rm T}\tilde \Theta =O$. The proximal mapping yields $\tilde \Theta = P_{A^{\perp}} B$, with $B\in\mathbb{R}^{p\times m}$ the matrix form for $\beta$. Using matrices $Y^{\rm T}\in \mathbb{R}^{n\times p}$, $X^{\rm T}\in \mathbb{R}^{n\times m}$, we rewrite the envelope regression likelihood as
\begin{align*}
& L(Y; A, W, B, \mu)  %\propto |\Omega|^{-n/2} \exp\bigg( -\frac{1}{2} \text{tr}  \bigg\{  [Y^{\rm T}- 1\mu^{\rm T}-  X^{\rm T} \theta^{\rm T} - Z^{\rm T} A^{\rm T}] [\Gamma_\theta   \Omega^{-1}           \Gamma_\theta^{\rm T}  ]    [Y  -\mu 1^{\rm T} -\theta X - A Z]   \bigg\}\bigg ) \\
 \\&  \propto
 \exp\bigg( -\frac{1}{2} \text{tr}  \bigg\{  [Y^{\rm T}- 1\mu^{\rm T}-  X^{\rm T} (P_{A^{\perp}} B)^{\rm T} ] (P_{A^{\perp}} W P_{A^{\perp}}^{\rm T}
    )^-    [Y  -\mu 1^{\rm T} - (P_{A^{\perp}} B) X ]
    \bigg\}\bigg )  \\
  & \times     |A^{\rm T}A|^{-n/2} \exp\bigg( -\frac{1}{2} \text{tr}  \bigg\{  [Y^{\rm T}- 1\mu^{\rm T}] (AA^{\rm T}
    )^{-}    [Y  -\mu 1^{\rm T} ] 
    \bigg\}\bigg ) |P_{A^{\perp}} W P_{A^{\perp}}^{\rm T}|_+^{-n/2}, \end{align*} 
where $|\cdot|_+$ is the pseudo-determinant.
% Using matrix form $Y^{\rm T}\in \mathbb{R}^{n\times m}$, $X^{\rm T}\in \mathbb{R}^{n\times p}$, we have the likelihood
% \begin{align*}
% L(Y; .) \propto |\Omega|^{-n/2} |\Omega_*|^{-n/2} \exp\bigg( -\frac{1}{2} \text{tr}  \bigg\{  [Y^{\rm T}- 1\mu^{\rm T}-  X^{\rm T} \theta^{\rm T}] [\Gamma \; \Gamma_*] 
%     \begin{bmatrix}
%         \Omega^{-1} & O\\
%         O &  \Omega_*^{-1} 
%     \end{bmatrix}
%         \begin{bmatrix}
%         \Gamma^{\rm T} \\
%          \Gamma_*^{\rm T} 
%     \end{bmatrix}
%     [Y  -\mu 1^{\rm T} -\theta X]
%     \bigg\}\bigg )
% \end{align*} 
To complete the proximal prior specification, we use $B\in \mathbb{R}^{p\times m}$ and  $R\in\mathbb{R}^{p\times p}$, with their elements iid from N$(0,1)$ (hence full rank almost surely); then we set $A= R \Lambda$, and $\Lambda = \text{diag}( \Lambda_{i,i})_{i=1}^p$ with $ \Lambda_{i,i} = z_{i} 1(z_i>\rho_q)$ and $z_i\sim \t{Exp}(1)$, and $\rho_q$ the $q$-quantile of $\t{Exp}(1)$. We provide a numerical simulation in the Supplementary Materials.

Focusing on the $\theta$-marginal prior, we have: 1. the mixture weight $\text{pr}^0[\text{rank}(A)=p-u] = {p \choose u} q^u (1-q)^{p-u}$, where $q$ can be specified as a priori; 2. the conditional density containing 
\begin{align*}
\Pi_{\bm\theta}^{0}\left(\theta \mid A\right) & =  (2\pi)^{-u/2} |P_{A^{\perp}}|_+^{-1/2} \exp( - \theta^{\rm T} P^{-}_{A^{\perp}} \theta /2 ) \\
& = (2\pi)^{-u/2} \exp( - \|\theta\|^2  /2 )1(A^{\rm T}\theta =0).
\end{align*}
The second line is due to $P_{A^{\perp}}$ being idempotent $\theta=P_{A^{\perp}}\beta=P_{A^{\perp}}\theta$, $P_{A^{\perp}}P^-_{A^{\perp}}P_{A^{\perp}}=P_{A^{\perp}}$, $\theta^{\rm T} P_{A^{\perp}} \theta = \theta^{\rm T}\theta$ and $|P_{A^{\perp}}|_+=1$. This density is invariant to scaling of $A$. 
 \end{example}

\begin{remark}
To clarify, we use the above example (with a relatively simple  $\Pi^0_{\bm \theta}$) to illustrate the equivalence between the  hierarchical  specification of $\Pi^0_{\bm \theta}$ and the continuous-$\Pi^0_{\bm \beta}$-and-mapping specification. In general cases, the former $\Pi^0_{\bm \theta}$ may be intractable due to the lack of closed-form, hence motivating the proximal mapping strategy as proposed in this article.
\end{remark}

In example 2, if we intuitively compare the two distributions  before and after the mapping, it reduces (or at least retains) the distance to the center $\| \theta- \t{prox}_{\lambda g}(\mu)\|_2 \le \|\beta -\mu\|_2$.
 The property shown in this example is known as  the ``non-expansiveness'', which in fact holds for all proximal mappings:
\[ \|\t{prox}_{\lambda g}(\beta_1)-\t{prox}_{\lambda g}(\beta_2)\|_2\le \|\beta_1-\beta_2\|_2,
\]
for any $\beta_1,\beta_2$ in the domain of $\t{prox}_{\lambda g}$. This is in particular meaningful for Bayesian inference, as it conveniently controls the concentration of measure for $\theta$.
\begin{theorem}
If the data $y$ come from a distribution $\mathcal F_{\theta^*}$ with a fixed parameter ${\theta^*}$, and any $\epsilon \in (0,1)$, the posterior distributions of $\theta$ and $\beta$ satisfy $$\pr(\|\theta-\theta^*\|>\epsilon\mid y) \le \int \min_{\beta^*: \t{prox}_{\lambda g_\gamma}(\beta^*)=\theta^*}\pr\left[\left\|\beta-\beta^{*}\right\|>\epsilon \mid y , \lambda, \gamma \right] \Pi(\lambda, \gamma\mid y) d(\lambda, \gamma).$$
In addition, if $\t{tr}[Cov(\beta\mid y)]<\infty$, then $$\t{tr}[Cov(\theta\mid y)]\le \t{tr}[Cov(\beta\mid y)].$$
\end{theorem}

Using the envelope regression example, we know that
\be
\Pi(B\mid y, A,W) \propto \exp\big\{  -(1/2)\t{tr}[ & XX^{\rm T}  B^{\rm T} P_{A^{\perp}} (P_{A^{\perp}} W P_{A^{\perp}}^{\rm T} )^- P_{A^{\perp}} B ] + \t{tr}(B^{\rm T}B)\\
&-2 \t{tr} [X  (Y^{\rm T}- 1\mu^{\rm T} ) (P_{A^{\perp}} W P_{A^{\perp}}^{\rm T}
    )^-   P_{A^{\perp}} B ])
\big \},
\ee
which is a multivariate Gaussian for $\t{vec}(B).$ On the other hand, since we know $\tilde\Theta= P_{A^{\perp}} B$, we know for any given $A$,
 $\|\tilde\Theta- \tilde\Theta^*\|_F^2 = \t{tr}[(B- B^*)^{\rm T}P_{A^{\perp}}P_{A^{\perp}} (B- B^*)]\le \|B- B^*\|_F^2$
  with $P_{A^{\perp}} B^*=\tilde\Theta^*$, due to $P_{A^{\perp}}$ being idempotent and  having eigenvalues equal to either $1$ or $0$.
 
% \bel\label{eq:concentration}
% \t{pr}(\|\theta- \mu^*\|_2 \le \epsilon)\ge \t{pr}(\|\beta- \mathbb{E}\beta\|_2 \le \epsilon),
% \eel
% for any $\epsilon>0$, where $\mu^*= \t{prox}_{\lambda g}(\mathbb{E}\beta)$. This  inequality holds for both the prior and the posterior distributions, hence is convenient for imposing prior assumption (such as when assuming $\theta$ to be near $\mu^*$ as a priori) and for the theoretic analysis of posterior concentration.
% %  (such as analyzing the concentration of $\beta$ as a convenient and unconstrained surrogate for $\theta$).
%  
% 
% 
% 
% 
% 
% \setcounter{example}{1}
% \begin{example}[(Continued). Affinely constrained prior]
% In the affinely constrained example,  
% if we choose the prior covariance of $\beta$ as $\Sigma=\sigma^2I$,  the prior has $\t{pr}[\| \theta-\t{prox}_{\lambda g}(\mu)\|_2\le  \sqrt{p} 2\sigma]\ge \t{pr}[\|\beta-\mu\|_2 \le \sqrt{p} 2\sigma] \ge 95\%$. Conveniently, this concentration holds for any rank of $A$.
% \end{example}

%\begin{remark}
%To clarify, we choose the above toy example to more intuitively illustrate the properties of the proximal prior. In general, we do not need the proximal mapping / prior to have a closed-form, yet all of the presented properties sill hold.
%\end{remark}

\subsection{\label{sec: prioronlambda}
Prior Specification on $\lambda$}
In the proximal mapping \eqref{eq:proximal_mapping}, the hyper-parameter $\lambda$ plays an important role, hence we need to carefully choose its prior. To first obtain some intuition, note when $\lambda \to 0$, we have $\prox(\beta)\to \beta$ if $g(z)<\infty$ for all $z$, the identity mapping; when $\lambda \to \infty$, we have $\prox(\beta)\to \argmin_zg(z)$. Therefore, as $\lambda$ increases, $\theta$ becomes farther away from $\beta$, hence the distribution  $\Pi_{\bm\beta}^0(\beta)$ gets more ``deformed'' at a larger $\lambda$.
 We now formalize this deformation intuition, while relaxing the finite-valuedness of $g$. For conciseness, we postpone all the proofs in the appendix.
\begin{theorem}[Monotonicity of deformation in $\lambda$] \label{thm1}
For any function $g$ with range $\mathbb{R}\cup\{\infty\}$, 
        if $0<\lambda_1<\lambda_2$, then $ \| \beta - \text{prox}_{\lambda_1 g} (\beta) \|_2\le \| \beta - \text{prox}_{\lambda_2 g} (\beta) \|_2$.
\end{theorem}

\begin{figure}[H]
        \centering
                \begin{subfigure}[t]{.45\textwidth}
        \includegraphics[height=4cm, width=5.5cm]{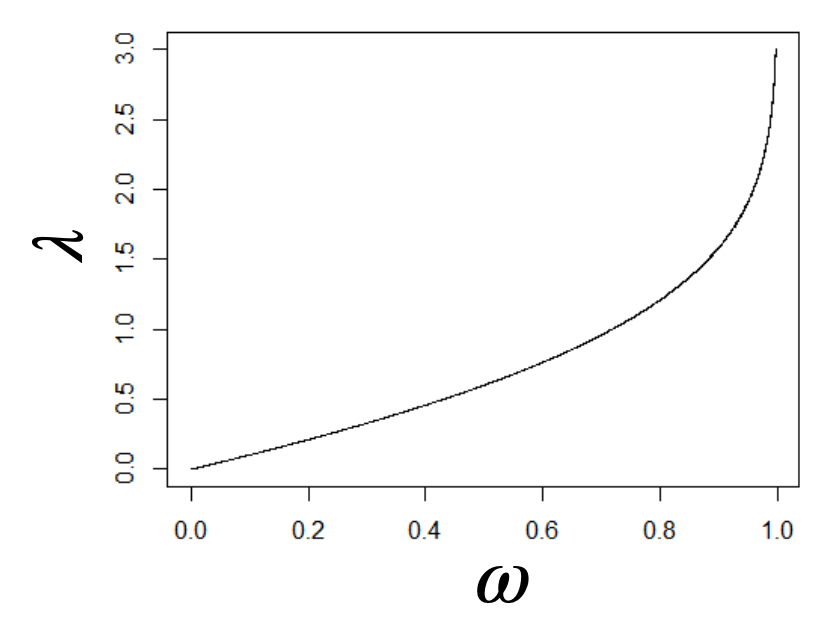}
        \caption{Value of $\lambda$ corresponding to a certain deformation $\omega$. Larger $\lambda$ leads to larger deformation from $\Pi_{\bm\beta}^0(\beta)$ to $\Pi_{\bm\theta}^0(\theta)$, which can be quantified by $\omega\in[0,1]$.}
        \end{subfigure}
        \qquad
                \begin{subfigure}[t]{.45\textwidth}
        \includegraphics[height=4cm, width=5.5cm]{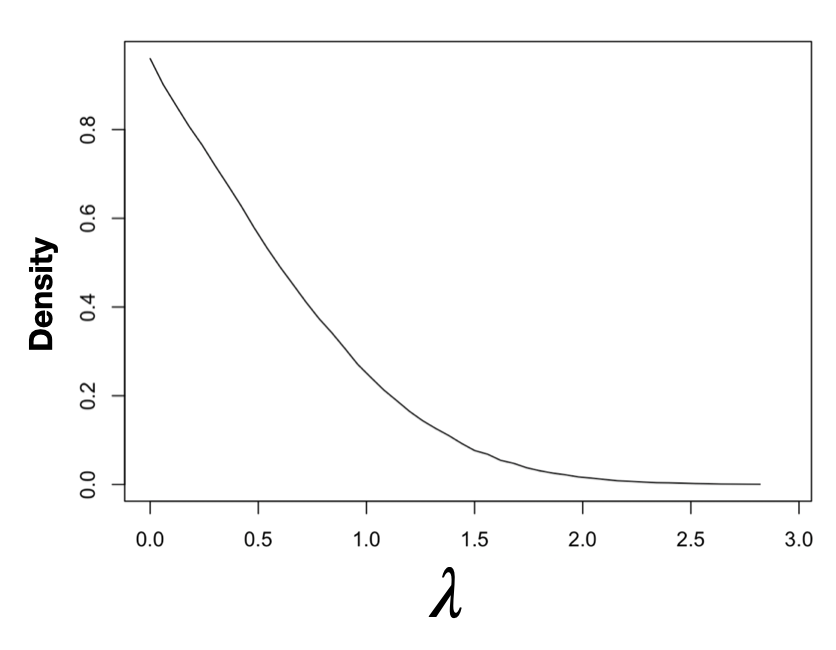}
                \caption{The prior for $\lambda$, corresponding to a uniform prior on the deformation measurement $\omega$.}
        \end{subfigure}
        \caption{\label{fig:lambda_2} Illustration of a numerically computed prior for $\lambda$, which controls sparse level for the soft-thresholding mapping $\theta=\t{sign}(\beta) \max(|\beta|-\lambda,0)$.}
\end{figure}

This result means that we can find a measurement between $0$ and $1$ to quantify the deformation: 
\bel
\omega_\lambda := \frac{ \mathbb{E}_{\bm\beta}\|\beta- \text{prox}_{\lambda g} (\beta) \|_2}
{\mathbb{E}_{\bm\beta} \| \beta - \lim_{\lambda^*\to \infty} \text{prox}_{\lambda^* g} (\beta)\|_2},
\eel
where the expectation is taken with respect to the prior of $\beta$.

When lacking prior knowledge on $\lambda$, we can use a Beta prior on $w\in(0,1)$ and solve for $\lambda$:
\bel
\label{eq: prior_lambda}
\omega \sim \text{Beta}(a_\omega,b_\omega), \qquad
\lambda = \min_{x>0} ( x: \omega_x = \omega).
\eel
In this article, we use a non-informative $a_\omega=b_\omega=1$. As a toy example, let $\beta$ be univariate with a finite variance, using the proximal mapping with $g(z) = z^2/2$, we have  $\prox(\beta) = {\beta}/({1+\lambda})$. Therefore, we have $\lambda=(1- \omega) / \omega$ with an induced prior $\Pi^0_{\bm\lambda}(\lambda) = 2/(1+\lambda)^2$ for $\lambda>0$.

In more general cases, \eqref{eq: prior_lambda} often cannot be solved analytically. However, we can numerically compute a prior for $\lambda$, using a  strategy similar to \cite{berger2009formal} --- for $K$ chosen points $\lambda_1,\ldots,\lambda_K$ in $(0,\infty)$, we can use the empirical estimates of the expectation based on  simulated   $\beta\sim \Pi^0_{\bm\beta}(\beta)$, and solve for $\omega_1,\ldots,\omega_K$; afterwards, we can easily interpolate to obtain the $\lambda$ associated with any $\omega$.

\setcounter{example}{0}
\begin{example}[(Continued). Soft-thresholding prior]
To illustrate, we compute the prior of $\lambda$ for the soft-thresholding prior based on $\theta=\t{sign}(\beta) \max(|\beta|-\lambda,0)$. Based on $\beta\in\mathbb R^p$ and $\beta\sim\No(0,I_p)$, we compute the prior density of $\lambda$ and plot it in Figure \ref{fig:lambda_2}.
\end{example}

In this section, we discussed the choice of $\Pi^0_{\bm \lambda}$  with the generality of all possible $g$ and $\prox$ in mind. On the other hand, for some specific case  such as $g(z)=\|z\|_1$ and soft-thresholding $\prox$, there is a connection to some existing prior in the literature, such as the classic spike-and-slab prior.
 For example, if $\Pi_{\bm \beta}^0 (\beta) \propto \exp(-\|\beta\|_1/\alpha) $ and $\lambda$ to be the $\omega$-quantile of of $\t{Exp}(\alpha^{-1})$, then we can obtain a spike-and-slab prior with Laplace slab $\Pi_{\bm \theta}^0(\theta \mid \lambda)= \prod_{j=1}^p[w_{\lambda} \delta_0(\theta_j) + (1-w_{\lambda})(2\alpha)^{-1} \exp(-\theta_j/\alpha)]$. 
A closely related discovery is the neuronized prior
\citep{shin2021neuronized} using truncated activation function, for which there is an
equivalence to  a spike-and-slab prior with two-normal-product slab. In these cases, there are often alternative choices for $\Pi_{\bm \lambda}^0$ that are justified via large sample theory.
Due to the page constraint, we defer the detailed discussion and numerical experiments to the Supplementary Materials.

\section{Geometric Measure Theory on the Varying Dimensional Sets}
\label{sec: theory}

\subsection{Hausdorff Dimension and Low Dimensional Density}
We now give a more rigorous exposition on the distribution induced by the proximal mapping. Without loss of generality, we consider $\theta$ as a $p$-element vector. Since $\theta$ may correspond to a measure of a set in the lower dimensional space, the $p$-dimensional Lebesgue measure of any lower-dimensional set is zero hence is not useful. We need some tools from the geometric measure theory to address this issue.
%A useful concept is the Hausdorff measure, which later will lead to the definition of the Hausdorff dimension.
 To start, consider a set $\mathcal A$ and suppose we do not know its dimensionality. Instead, we can cover $\mathcal A$ with sets ${B}_i$'s, each $B_i$ has its diameter $\text{diam}(B_i)=\sup\{|x-y|:x,y\in B_i\}\le\delta$. We call any $\bigcup_i B_i\supset \mathcal A, \text{diam}(B_i)<\delta$ as a $\delta$-covering of $\mathcal A$.

Then we take the infimum over all  $\delta$-coverings of $\mathcal A$, and letting the $\delta$ decrease, we obtain the  $s$-dimensional Hausdorff measure of $\mathcal A$:
\bel
\mathcal H^s(\mathcal A)=\lim_{\delta\to 0}\inf \left\{  \sum_{i=1} ^{\infty} \t{diam}(B_i)^s: {\text{diam}(B_i)\le\delta, \mathcal A\subseteq\bigcup_i B_i} \right\}.
\eel
Intuitively, the above can be taken as the minimum total ``volume'' of the covering --- except $s$ is a parameter that varies.

In fact, $\mathcal H^s(\mathcal A)$ is a non-increasing function in $s\ge 0$ \citep{edgar2007measure}. More importantly, for any Borel $\mathcal A$, and $0<s_1<s_2$, if $\mathcal H^{s_1}(\mathcal A)<\infty$ then $\mathcal H^{s_2}(\mathcal A)=0$; and if $\mathcal H^{s_2}(\mathcal A)>0$ then $\mathcal H^{s_1}(\mathcal A)=\infty$ [Theorem 6.1.6 \citep{edgar2007measure}]. This means for any Borel set $\mathcal A$, there is a unique $s_0\in [0,\infty) \cup \{\infty\}$ as a transition point, over which the dimensionality drops from $\infty$ to $0$:
\be
 \mathcal H^s(\mathcal A) =\infty, &\text{ for any } s<s_0; \\
 \mathcal H^s(\mathcal A) =0, &\text{ for any } s>s_0. \\
\ee
Such an $s_0$ is referred to as the Hausdorff dimension of $\mathcal A$, equivalently:
 \bel
 \dim_{\mathcal H}(\mathcal A) =\inf\{s\ge 0:\mathcal H^s(\mathcal A)=0\}.
 \eel
Note that $ \dim_{\mathcal H}(\mathcal A)\ge 0$ does not have to be an integer; nevertheless, when it is, the Hausdorff measure is proportional to the commonly used $s$-dimensional Lebesgue measure
\be
\lambda^{s}(\mathcal A)= \inf \left\{ \sum_{i=1}^\infty \t{vol} (B_i): \mathcal A\in \bigcup B_i, B_i \t{ is an open cube} \right\},
\ee
via $\lambda^{s}(\mathcal A) = w_s \mathcal H^s(\mathcal A)$, where $w_s=\pi^{s/2} [2^s \Gamma(s/2+1)]^{-1}$ due to the volume formula of an $s$-dimensional ball. In addition, when $s=0$, $\mathcal H^0(\mathcal A)$ is same as the counting measure.

Now recall that $\prox$ is non-expansive, which leads to the following theorem:
  \begin{theorem} \label{thm2}
  For any Borel set $\mathcal A$ and proximal mapping $\prox$, we have
  \begin{enumerate}
  \item $\mathcal H^s\{\prox(\mathcal A)\} \le  \mathcal H^s(\mathcal A) $ for any $s\ge 0$;
  \item  $\dim_{\mathcal H}\{\prox(\mathcal A)\}\le \dim_{\mathcal H}(\mathcal A)$.
    \end{enumerate}
  \end{theorem}

\begin{remark}
In the above, the statement 2 is particularly useful: it tells us that $\prox$ only maps to lower or equal dimensional space.
\end{remark}

Now, starting from a probability distribution defined by a certain Radon measure $\mu$ in $\mathbb{R}^p$ for some low-dimensional sets in $\Theta^s$, one interesting question is how to differentiate this and obtain a ``density'', as $\Pi_{\bm \theta}(\theta=t\mid \theta \in \Theta^s)$ used in \eqref{eq:prior_kernel} and \eqref{eq:posterior_kernel}.
  
For a point $\theta\in\Theta^s$, the ball  $B_r(\theta)$ centered at $\theta$ with radius $r>0$ has the  lower and upper $s$-dimensional derivatives:
\be
f^s_{\mu,*}(\theta)=\lim\inf_{r\to 0} \frac{\mu\{ B_r(\theta)\}}{w_s r^s}, \qquad f_{\mu}^{s,*}(\theta)=\lim\sup_{r\to 0} \frac{\mu\{ B_r(\theta)\}}{w_sr^s}.
\ee
Therefore, if we have the two limits coincide, we would have a definition of an $s$-dimensional density:
$f^s_{\mu}(\theta) = f^s_{\mu,*}(\theta)= f^{s,*}_{\mu}(\theta)$, commonly referred to as the $s$-density.

\begin{remark}
        To understand the $s$-density as a generalized concept of ``density'', for those continuous distributions associated with a $p$-dimensional Lebesgue measure, such as the non-degenerate Gaussian distribution, the $p$-density is the probability density function; whereas for the discrete distributions, the $0$-density is the same as the probability mass function.
\end{remark}

Next, similar to the probability density function, $s$-density may not always exist. Therefore, it is important to state the two required conditions, $s$ is an integer  and $\theta$ is in a rectifiable set, as formalized in the following theorem.
\begin{theorem}[Besicovitch-Marstrand-Preiss theorem] 
\label{density_existence_theorem}
\citep{preiss1987geometry}
Let $\mu$ be a locally finite Radon measure on $\mathbb{R}^{p}$, if there exists a real $s\ge 0$ such that $f^s_{\mu}(\theta)$ exists, and it is positive on a set of positive $\mu$-measure, then $s$ must be an integer. On the other hand, let $\mathcal A \subset \mathbb{R}^{p}$ be Borel with $\mathcal H^s(\mathcal A)\in(0,\infty)$ and $s$ an integer, then $f^s_{\mu}(\theta)$ exists for $\theta \in \mathcal A$ almost everywhere with respect to $\mathcal{H}^{k}$, if and only if the set $\mathcal A$ is rectifiable.
\end{theorem}
To explain ``rectifiability'', a Borel set $\mathcal A \subset \mathbb{R}^{p}$ is rectifiable if 
%$\t{dim}_\mathcal H(\mathcal A)=s$ and 
there is a countable family of Lipschitz maps $T_i: \mathbb{R}^{s}\to \mathbb{R}^p$ which cover almost all $\mathcal A$ except for sets with zero $\mathcal{H}^{s}$ measure. That is, intuitively speaking, almost every $p$-element vector $\theta\in \mathcal A$ can be represented as some transformation of $x\in\mathbb{R}^s$ --- note that this is not the same as a simple reparameterization, as we may obtain $\mathcal A$ via multiple $f_i$'s (up to countably many).

%\maoran{Though it seems hard to determine the existence of the $s$-density while the dimensionality of $\Theta$ is unknown or varying, in practice, fortunately, often the dimensionality is more straightforward to find out. In this article, all of the presented proximal mappings have valid $s$-densities almost surely with respect to the prior / posterior distribution. For example, the image of soft-thresholding with $k$ non-zeros has Hausdorff dimension $k$; similarly, the norm-constrained space, the low-rank matrix space and the positive-definite matrix space all have integer Hausdorff dimension.} 

\subsection{Calculation of the $s$-Density}
We now provide a way to calculate the $s$-density. Focusing on a subset $\Theta^k$ with $\t{dim}(\Theta^k)=k$ and $\Beta^k = \prox^{-1}(\Theta^k)$. We now transform $\Pi_0(\beta=b \mid \beta\in \Beta^k)$ into an $s$-density with $s=k$.

\begin{theorem} \label{thm4}
If $\Beta^k$ is $(\mathcal H^{p},p)$-rectifiable and $\t{dim}_{\mathcal H}(\Beta^k)=p$, $\Theta^k$ is $(\mathcal H^{k},k)$-rectifiable and $\t{dim}_{\mathcal H}(\Theta^k)=k$, with $p\ge k$, and  $J_{k} \prox(\beta)>0$ a.e.-$\mu_\beta$. Then the $s$-density of $\theta$ induced by $\prox$ is
\bel\label{eq:transformation_formula}
\Pi(\theta = t\mid \theta\in \Theta^k)= \int_{\prox^{-1}(t)}\frac{\Pi(\beta =b)/ \textup{pr} \{\prox(\beta)\in \Theta^k\}}{J_{k} \prox(b)} w_{(p-k)} d\mathcal H^{p-k}(b),
\eel
where $J_{k} \prox(b)$ is the $k$-dimensional Jacobian of $\prox$ at b.
\end{theorem}
Note that if the low-dimensional set $\Theta^k$ can be reparameterized as a transformation an $k$-element vector, then it is possible to change \eqref{eq:transformation_formula} to an integration with respect to an $k$-dimensional Lebesgue measure.

%\begin{remark}
%This theorem applies to both prior $\Pi_0^{\theta}( t\mid \theta\in \Theta^k)$ and posterior $\Pi(\theta = t\mid \theta\in \Theta^k, y)$.
%\end{remark}

To explain the assumptions above, a set $\mathcal A$ is $(\mathcal H^{s},s)$-rectifiable when $\mathcal H^{s}(\mathcal A)<\infty$, 
and there is a set as the countable union of Lipschitz images from bounded sets $\mathcal B = \bigcup_j \{ T_j(\mathcal C_j): \mathcal C_i\subset \mathbb{R}^s \t{ and bounded}, T_j \t{ Lipschitz}\} $ such that $\mathcal H^{s} (\mathcal A\setminus \mathcal B)=0$. As the result, if $s$-density exists, we could use \eqref{eq:transformation_formula} when both $\theta$ and $\beta$ are finite.

\cite{morgan2016geometric} gives the $k$-dimensional Jacobian $J_k T(x)$ of function $T: \mathbb R^n\to\mathbb R^m$, differentiable at $x$. Let $D_T(x)\in\mathbb R^{n\times m}$ be the derivative matrix of $T(x)$ at $x$, with 
$\{D_T(x)\}_{ij}={\partial T(x)_j}/{\partial x_i}$, $1\le  i\le n, 1 \le j\le m$, then the $k$-dimensional Jacobian can be computed as:
\bel
J_k T(x)=\sqrt{\sum_{\substack{ M \t{ is the $k\times k$} \\ \t{submatrix of }  D_T(x)}} ( \det  M )^2}.
\eel
Note than when $m=n=k$, $J_k T(x)=| \det\{D_T(x)\}|$ as more commonly seen.

Importantly, by  Rademacher's theorem \citep{federer2014geometric}, a Lipschitz function is differentiable almost everywhere. Therefore $D_{\prox}(\beta)$ exists almost surely with respect to $\mu_\beta$. In the following example, we illustrate the use of the above theorem to compute the $s$-density for the affinely constrained prior.

\setcounter{example}{1}
\begin{example}[The $s$-density of the affinely constrained prior]
Using \eqref{eq:transformation_formula}, one can verify that the $s$-density of affinely constrained prior recovers the ``degenerate Gaussian density''. Starting from $\beta \sim  \No(\mu,\Sigma)$ and let us assume $A$ is $p\times d$ and $\t{rank}(A)=d$, then it is not hard to compute that $J_{(p-d)} \prox(\beta) = 1$. Using the proximal mapping, at $\theta=t$, we have $P_A\beta =t - A(A^\T A)^{-1}b $ with $P_A={\{I-A(A^\T A)^{-1}A^\T\}}$,  hence we can integrate over the region $\prox^{-1}(t)$ by re-parameterizing $\beta= P_A^{-} \{t - A(A^\T A)^{-1}b \} + A x$ where $x\in\mathbb{R}^d$. Integrating over $x$, we have the $s$-density with $s=p-d$:
\be
\Pi_{\bm\theta}^0(t \mid \theta\in \Theta^{p-d}) \propto \exp \left[ - \frac{1}{2} \{t -  A(A^\T A)^{-1}b -P_A\mu\}^\T
P_A^{-}\Sigma^{-1} P_A^{-} \{t -  A(A^\T A)^{-1}b -P_A\mu\} \right],
\ee
which is commonly referred to as the ``degenerate density'' for a degenerate Gaussian, with its covariance $P_A\Sigma P_A$ having a rank $(p-d)$.
\end{example}
We list a few more examples commonly considered in statistics, where for each we have a guaranteed existence of $s$-density: regression under linear equality constraints, matrix factorization under low-rank constraint, sparse regression, covariance modeling in positive-definite space, and directional modeling in orthonormal space.

\begin{remark} 
To clarify, the existence of $s$-density for $\theta$ is not necessary in our modeling framework using proximal mapping, since we can always carry out computation using a valid $p$-dimensional density of $\beta$. On the other hand, the existence of $s$-density would be required if one wants to interpret the prior via an equivalent prior $\Pi^0_{\bm \theta}$ as  in \eqref{eq:prior_kernel}.
 \end{remark}

%\begin{remark}
%\end{remark}

%\begin{remark}
%With Theorem \ref{thm4}, we can now use Equation \ref{eq:posterior_kernel} to compute the posterior of $\theta$.  \maoran{However, practically, we consider the prior selection probability $\text{pr}(\theta\in \Theta^k)$ to be implicitly induced via the choice of $\Pi^0_{\bm \beta}$ and proximal mapping. This is motivated by our applications where we do not have any strong prior belief on dimensionality. On the other hand, when there is a need for a ``balanced'' prior distribution on the dimensions, one can calibrate the choice of $\Pi^0_{\bm \beta}$ by numerically examine the empircal prior probability $\text{pr}(\theta \in \Theta^k)$ denoted by $v_k$, and then replace a reweighted prior distribution $\tilde \theta^0_{\bm \beta}(b) = v_k/(\sum_{k'} v_{k'}) \Pi^0_{\bm \beta}(b) 1[\text{prox}(b)\in \Theta^k]$. }
%\end{remark}
%Although less useful, for the sake of theoretic interest, at the point of non-differentiability, the Jacobian can be replaced by the approximate Jacobian and \eqref{eq:transformation_formula} still holds \citep{federer2014geometric}.

 \section{ \label{sec: posterior}Posterior Computation}
As shown in (\ref{eq:post_density_beta}), when using $\beta$ instead of $\theta$, the posterior has a simple density on $\mathbb R^p$, and the proximal mapping is differentiable almost everywhere with respect to $\mu_\beta$. Therefore, as long as $\Pi(\theta;y)$ is a continuous and differentiable function in $\theta$ almost everywhere, we can use the  Hamiltonian Monte Carlo (HMC) for posterior computation. Now we first briefly review the HMC algorithm, then address the gradient calculation for the proximal mapping. 
 
 To sample from target distribution $\beta\sim\Pi_{\beta\mid y}(\cdot)$, the  HMC uses an auxiliary momentum variable $v$ and samples from a joint distribution $\Pi(\beta,v)=\Pi(\beta\mid y)\Pi(v)$, where a common choice of $\Pi(v)$ is the density of $\No(0,M)$. Denote $U(\beta) = -\log\Pi(\beta\mid y)$ and $K(v) = -\log\Pi(v)=v^\T M^{-1}v/2$, which are referred to as the potential energy and kinetic energy respectively. The total Hamiltonian energy function is $H(\beta,v) = U(\beta)+K(v)$. 
 
 At each state $(\beta,v)$, a  new state is generated by simulating Hamiltonian dynamics, which satisfies the Hamilton's equations:
  \bel
  \label{eq: diff_equation}
  \frac{\partial \beta}{\partial t} = \frac{\partial H(\beta,v)}{\partial v} = M^{-1}v;\quad \frac{\partial v}{\partial t} = -\frac{\partial H(\beta,v)}{\partial \beta} = \frac{\partial\log\Pi(\beta\mid y)}{\partial\beta}.
 \eel 
The exact solution for (\ref{eq: diff_equation}) is often intractable, while we can numerically approximate
the evolution  by algorithms such as the leapfrog scheme. The leapfrog is a reversible and volume-preserving integrator, which updates the evolution $(\beta^{t},v^{t})\to (\beta^{t+\epsilon},v^{t+\epsilon})$ via
\bel
v\leftarrow v+\frac{\epsilon}{2} \frac{\partial\log\Pi(\beta\mid y)}{\partial\beta},\quad \beta\leftarrow \beta+\epsilon M^{-1}v,\quad v\leftarrow v+\frac{\epsilon}{2} \frac{\partial\log\Pi(\beta\mid y)}{\partial\beta}
\eel 
for $t=0,\epsilon,\ldots, L\epsilon$, and sets $(\beta^*,v^*)\leftarrow(\beta^{L\epsilon},v^{L\epsilon})$. To correct the numeric error due to approximation, $(\beta^*,v^*)$ is treated as a proposal and  accepted with the Metropolis-Hastings (MH) probability
 \be
 \min [1, \exp\{-H(\beta^*,v^*)+H(\beta,v)\}].
 \ee

We now discuss the gradient computation:
\be
\frac{\partial\log\Pi(\beta\mid y)}{\partial\beta} = 
\frac{\partial \prox (\beta)}{\partial \beta} 
\left\{ \frac{\partial \log L( y;\theta)}{\partial \theta}\bigg\vert_{\theta=\prox(\beta)}\right\}
+ \frac{\partial \log\Pi^0_{\bm\beta}(\beta)}{\partial\beta}.
\ee
When $\prox(\beta)$ has a closed-form, we can use the automatic differention toolbox to calculate the gradient ${\partial \prox (\beta)}/{\partial \beta} $; on the other hand, when the closed-form does not exist, some numeric approximation is needed.

% For numeric gradient of ${\partial \prox (\beta)}/{\partial \beta} \in \mathbb{R}^{p\times p}$,  one would compute $\{\prox(\beta+ \epsilon e_k)-\prox(\beta)\}/\epsilon$ for $p$ times, with $e_k\in \{0,1\}^p$ a vector all equal to zero except for the $k$the element equal to one, $\epsilon>0$ some small number. However, when $\beta \in \mathbb{R}^p$ is in high dimension, this would lead to a major computational burden.

% To solve this problem, we use the directional gradient. Suppose $\dot{\theta}={\partial \log L( y;\theta)}/{\partial \theta}$ is tractable and can be efficiently computed. Then we have
% \begin{eqnarray*}
% \frac{\partial \prox (\beta)}{\partial \beta_k} \dot{\theta}=   \lim_{\epsilon\to 0} \frac{\prox (\beta + {\dot{\theta}} \epsilon)
% -  \prox (\beta )}{\epsilon},
% \end{eqnarray*}
% which can be approximated using a small $\epsilon$ and evaluating $\prox (\beta + {\dot{\theta}} \epsilon)$ for one time at each leap-frog step.

% For example, one could use relationship between the proximal mapping the Moreau envelop, while relying on the second derivative of the latter for the calculation of ${\partial \prox (\beta)}/{\partial \beta} $ \{for the details, see \citep{poliquin1996generalized}\}.%
 Note that the partial gradient is  ${\partial \prox (\beta)}/{\partial \beta_j}=\lim_{\epsilon\to 0} \{ \prox(\beta+e_{j }\epsilon) -\prox(\beta) \}/{\epsilon}$ with $e_j$ is the standard basis with the $j$th element equal to one, and all others equal to zero; using a small $\epsilon$ gives us the finite difference approximation. Nevertheless, when $\beta$ is high dimensional, this would involve $(p+1)$ times of calculating the proximal mapping, which can be computationally prohibitive. To solve this problem, we follow \cite{spall1992multivariate} and use the simultaneous perturbation stochastic approximation:
\bel\label{eq: set_proj}
\frac{\partial \prox (\beta)}{\partial \beta_j} \approx \frac{1}{m} \sum_{k=1}^m\frac{\{ \prox(\ \beta+ \Delta^{(k)}\epsilon) -\prox(\beta) \}}{\Delta^{(k)}_{j }\epsilon},
\eel
for $j=1,\ldots,p$, 
where $\Delta^{(k)}=\{\Delta^{(k)}_1,\ldots, \Delta^{(k)}_p\}$ has each $\Delta^{(k)}_j\in\{-1,1\}$ independently generated using $\pr\{\Delta^{(k)}_j=1\}=\pr\{\Delta^{(k)}_j=-1\}=0.5$. The right hand side is based on the first order
approximation to the finite difference form. The advantage is that we only need to evaluate the proximal mapping for $m$ times.
 In this article, we use $\epsilon=10^{-7}$ and $m=20$ and find empirically good  stability for the HMC algorithm. 

For the HMC as a gradient-based algorithm, another potential concern is that $\prox(\beta)$ may have zero gradient at certain value of $\beta$, for example, the soft-thresholding $\t{sign}(\beta) \max(|\beta|-\lambda,0)$ will have zero gradient for those $\beta_j:|\beta_j|<\lambda$. Fortunately, two things prevent such a $\beta_j$ from  being stuck at a certain value. First, although the log-likelihood $\log L[y;\prox(\beta)]$ may have a zero gradient for $\beta_j$, the log-prior $\log\Pi^0_{\bm \beta}$ does not (as it does not depend on $\prox$) --- in those cases,   $\beta_j$ will be updated through its prior distribution, until it enters the region where $L[y;\prox(\beta)]$ is no longer invariant in $\beta_j$. This behavior is quite similar to the one with augmented ``continuous particle'' for sampling binary distribution via HMC \citep{pakman2013auxiliary}, where they demonstrated excellent mixing of Markov chains.  Second, the HMC preserves the joint density of $\Pi(\beta \mid y) \Pi(v)= \Pi(\beta^* \mid y) \Pi(v^*)$ (with the MH correction), and as we sample a new $v$ at the start of each iteration, the effective range of $\beta^*$ to reach is $\{ \beta^*:\Pi(\beta^*\mid y)= \Pi(\beta \mid y) \Pi(v)/\Pi(v^*),  \Pi(v^*)>0, \Pi(v^*)\le\Pi(\hat v)\}$, with $\hat v$ the mode of $\Pi(v)$. Therefore, as long as $\Pi(v)< \Pi(\hat v)$, we can have  $\Pi(v^*)> \Pi(v)$, and $\Pi(\beta^*\mid y)< \Pi(\beta\mid y)$ allowing $\beta^*$ to move away from a local-optimal state.
In practice, we use the No-U-Turn algorithm \citep{hoffman2014no}, which ensures that we run the dynamics for long enough, so that the new proposal is away from the current sate. We provide some diagnostic plots in the Supplementary Materials.

\section{Simulation Studies}
\subsection{Set Expansion Prior for Hypothesis Testing}
We now demonstrate the usefulness of the proximal prior in standard statistical inference, such as the hypothesis testing of whether $\theta$ is in a constrained set $C$. Consider two hypotheses $H_0: \theta\in C$ and $H_1: \theta\in\bar C$, where $C\bigcap\bar C = \varnothing.$ For testing, one typically assumes a mixture prior
\bel
\label{eq: hypo_prior}
\Pi^0_{\bm\theta}(\theta)= p_0 \phi(\theta) {\bf 1}({\theta\in C}) + (1-p_0) \bar\phi(\theta) {\bf 1}({\theta\in \bar C}),
\eel
where $\phi$ and $\bar\phi$ are the prior kernel function of $\theta$ under $H_0$ and $H_1$, respectively; and $p_0$ is the prior probability assigned to $C$. The Bayes factor of $H_0$ relative to $H_1$ is defined as 
\be
\operatorname{BF_{01}} = \frac{\int_{C}L(y ; \theta)\phi(\theta)\textup{d} \theta }{\int_{\bar C}L(y ; \theta)\bar\phi(\theta)\textup{d}\theta}
=\frac{\pr(\theta \in C\mid y)}{\pr(\theta \in \bar C \mid y)}  \frac{(1-p_0)}{p_0},
\ee
for which, a smaller value of $\operatorname{BF_{01}} $ provides stronger evidence against $H_0$. Often, $C$ is not of the same dimension with $\bar C$. For example, when testing a point null hypothesis $H_0:\theta_1 = 0$, we have $\operatorname{dim}(C) <\operatorname{dim}(\bar C)$. The standard practice has been assigning appropriate $\phi$  under $C$ (and $\bar \phi$ under $\bar C$), with $\phi(\theta)$ and $\bar\phi(\theta)$ being 0 on $\bar C$ and $C$ respectively.  However, when the null hypothesis is low-dimensional, such as testing linear equality, assigning density supported on the null set $C$ can become quite challenging.
\begin{figure}[H]
    \centering
         \begin{subfigure}[t]{.48\textwidth}
        \centering
        \includegraphics[trim={2cm 2cm 2cm 2cm},clip,width = 1.1\textwidth]{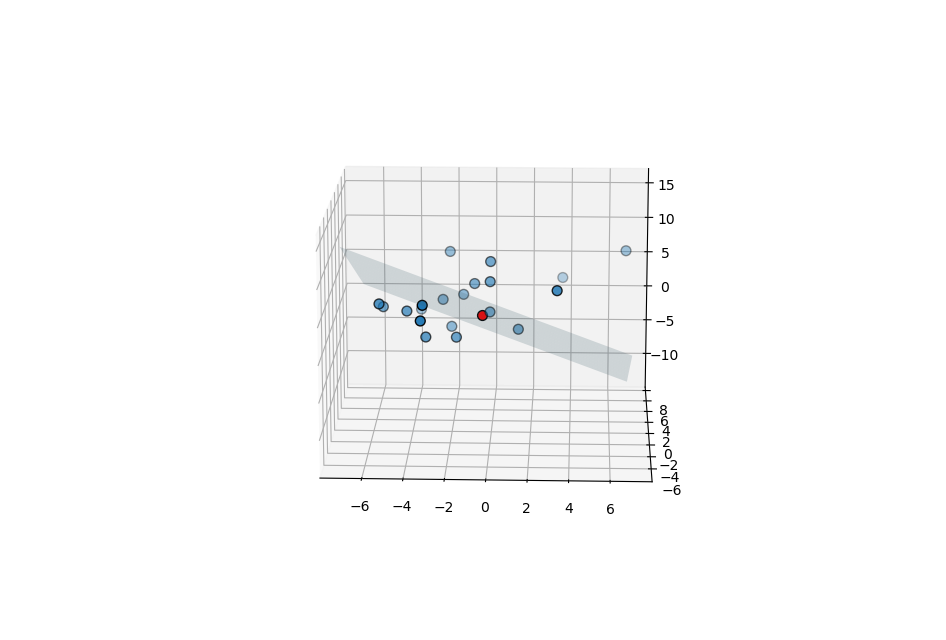}
        \caption{The data $y_1,\ldots,y_{20}\sim \No(\theta_0, 3^2I_3)$ (the blue dots), with the true mean $\theta_0=(-0.5,0.3,1.2)^\textrm{T}$ (the red dot), shown along with the hyperplane $C=(\theta: \theta_1 + \theta_2 + \theta_3 = 1)$. }
\end{subfigure}
\quad
        \begin{subfigure}[t]{.48\textwidth}
         \centering
         \includegraphics[trim={2cm 2cm 2cm 2cm},clip,width = 1.1\textwidth]{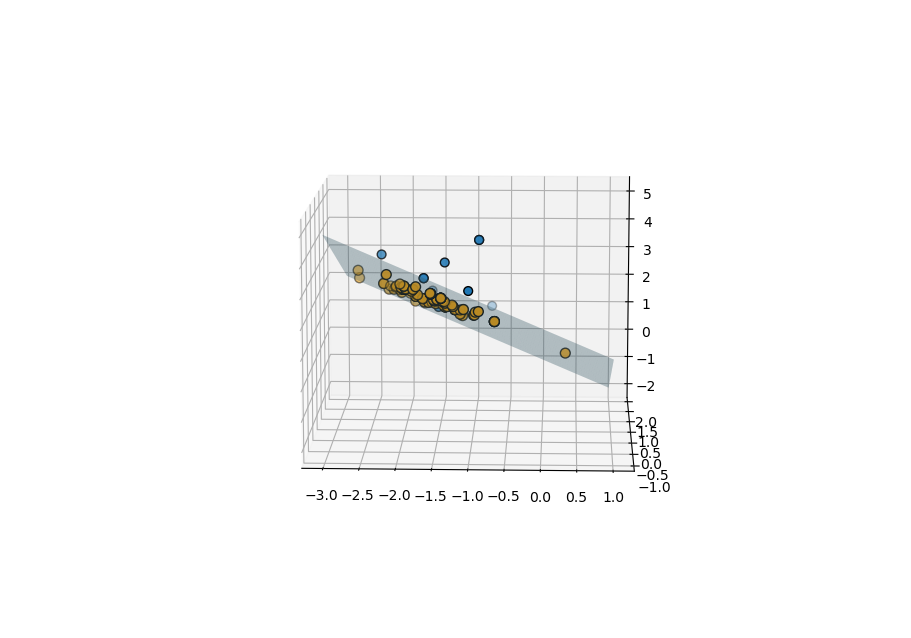}
        \caption{The posterior samples from $\pi(\theta\mid y)$, which are mostly distributed on the hyperplane (the orange dots), while there are several outliers far from this hyperplane (the blue dots).}
\end{subfigure}
    \caption{The set expansion prior for testing  $\{\theta=(\theta_1,\theta_2,\theta_3): \theta_1 + \theta_2 + \theta_3 = 1\}$.}
    \label{fig: set_exp}
\end{figure}
For a convex null set $C$, we could define a proximal prior based on the distance function, such that the prior density is positive on both $C$ and $\bar C$. The distance function from point $\beta$ to set $C$ is defined as $ \t{dist}_{C}(\beta) = \inf_{x\in C}\|x-\beta\|_2 = \|\beta-P_{C}(\beta)\|_2$.
The proximal mapping of the distance function to set $C$ is of the form
% The Euclidean distance from a point $\beta$ to $C$ is
% \be 
% \t{dist}_{C}(\beta) = \inf_{x\in C}\|x-\beta\|_2 = \|x-P_{C}(x)\|_2,
% \ee 
% which is a convex and lower semi-continuous function, and $P_{C}$ denotes the Euclidean projection onto ${C}$. The proximal mapping based on the distance function is 
\be
\t{prox}_{\lambda\t{dist}_{C}}(\beta) = \begin{cases}\beta + \frac{\lambda}{\t{dist}_{C}(\beta)}\{ P_{C}(\beta) - \beta \}, & \t{if dist}_{C}(\beta) \ge \lambda\\
P_{C}(\beta), & \t{if dist}_{C}(\beta) < \lambda.
\end{cases}
\ee 
 Clearly, this proximal mapping projects the points in the $\lambda$-neighborhood of $C$ into $C$, and keeps the rest of the points out of $C$. Thus
 we get a prior that puts positive mass on both $C$ and $\bar C$ and can also be expressed in the form of \eqref{eq: hypo_prior}.

We can easily estimate the Bayes factor 
\be
\operatorname{BF_{01}} = 
\frac{\pr \{ \t{dist}_{C}(\beta) < \lambda \mid y\}}{\pr \{ \t{dist}_{C}(\beta) \ge \lambda \mid y\}}
  \frac{\pr \{ \t{dist}_{C}(\beta) \ge \lambda\}}{\pr \{ \t{dist}_{C}(\beta) < \lambda\}}
\ee
via posterior sampling methods. If the prior ratio $\pr(\theta\in C)/\pr(\theta\in\bar C)$ is not specified, in order to obtain adequate number of samples in both $C$ and $\bar C$, we can choose a fixed $\lambda$ (instead of assigning a prior on $\lambda$) such that $\pr \{ \t{dist}_{C}(\beta) \ge \lambda \} \approx \pr \{ \t{dist}_{C}(\beta) < \lambda\}$.

We conduct a simulated experiment: we have data $y_1,\ldots,y_{20}$ generated from $\No(\theta_0,3^2)$ with $\theta_0 = (-0.5,0.3,1.2)^\T$. We want to test the linear equality hypothesis $H_0: \theta_1+\theta_2+\theta_3 = 1$ against $H_1: \theta_1+\theta_2+\theta_3 \neq 1$. The null set $C$ is a hyperplane with Hausdorff dimension 2. We assign the set expansion prior to $\theta$ by assign $\No(0, 3^2)$ to $\beta$ and set $\theta = \t{prox}_{\lambda\t{dist}_{C}}(\beta)$ with $\lambda = 2$, such that the prior ratio of $C$ and $\bar C$ is around $0.48.$ Posterior sampling is implemented with the HMC with 5000 samples and 2000 burn-ins. We get an estimated Bayes factor $BF_{01} = 0.77$, and display 100 of the samples in Figure~\ref{fig: set_exp}, panel (b).

\subsection{Numerical Experiments on Variable Selection and Low Rank Matrix Model}
In addition, we conduct numerical experiments for two models where solutions exist with conventional sparse priors: variable selection using a spike-and-slab prior, and low-rank matrix factorization with a discrete prior on the rank. We compare the computational performance in the combinatorial search-based MCMC algorithms for these models, with the HMC algorithm for our models using proximal priors. Further, we compare with other alternatives such as neuronized prior \citep{shin2021neuronized} and multiplicative shrinkage prior \citep{bhattacharya2011sparse,legramanti2020bayesian}. We provide the details in the Supplementary Materials.

 \section{Data Application: Interpretable Factor Analysis of the Flow Network}
 
    We now demonstrate the practical usefulness of the proximal prior via  analyzing the dynamic flow network data. The data \citep{zhu2020estimating} include dynamic estimated traffic flow on major roads in Florida every 6 hours before Hurricane Irma made landfall until it covered the entire state (between 18:00 on September/6/2017 and 18:00 on September/11/2017). In total, the data contain 25 valid temporal records of flow networks, denoted by $Y^{(1)},\ldots,Y^{(25)}$; each $Y^{(t)} \in \mathbb{R}^{n_V\times n_V}$ contains the traffic flows during a 6-hour period on the roads between $n_V=382$ urban regions.
    
     Each flow network is a weighted graph $\{V,E, Y^{(t)}\}$, with $V=(1,\ldots, n_V)$ the set of $n_V$ nodes, $E = \{(i, j): i,j\in V\}$ the edges, and the weight $Y_{i,j}^{(t)}\in\mathbb R$, representing the amount of flow between the two nodes, with $Y^{(t)}_{i,j}>0$ a flow $i\to j$, and $Y^{(t)}_{i,j}<0$ a flow $j\to i$. On the diagonal, $Y^{(t)}_{i,i}>0$ indicates an external in-flow entering the network, while $Y^{(t)}_{i,i}<0$ means an exiting out-flow; $Y^{(t)}_{i,j}=0$ if $(i,j)\not\in E$.

 To find useful patterns underneath the raw observation data, we use a low-dimensional latent factor model, with the factors $F^{(l)}\in \mathbb{R}^{n_V\times n_V}$ shared by all time points, while letting the loadings $\gamma_l^{(t)} \ge 0$ vary over time, subject to Gaussian measurement error $\mathcal E^{(t)}_{i,j}\stackrel{\text{iid}}\sim \No(0,\sigma^2_{\mathcal E})$ for $i\le j$.
      \bel
  \label{eq: factor_model}
  Y^{(t)} = \sum_{l=1}^d \gamma_l^{(t)} F^{(l)} + \mathcal E^{(t)}.
  \eel 
  \vspace{-1.cm}
\begin{figure}[htbp]
\centering
     \begin{subfigure}[t]{.3\textwidth}
\centering
        % \vspace{.3cm}
        \setlength{\fboxrule}{.1pt}
        \setlength{\fboxsep}{-.1cm}
        \fbox{\includegraphics[width=4.4cm, height=3.3cm]{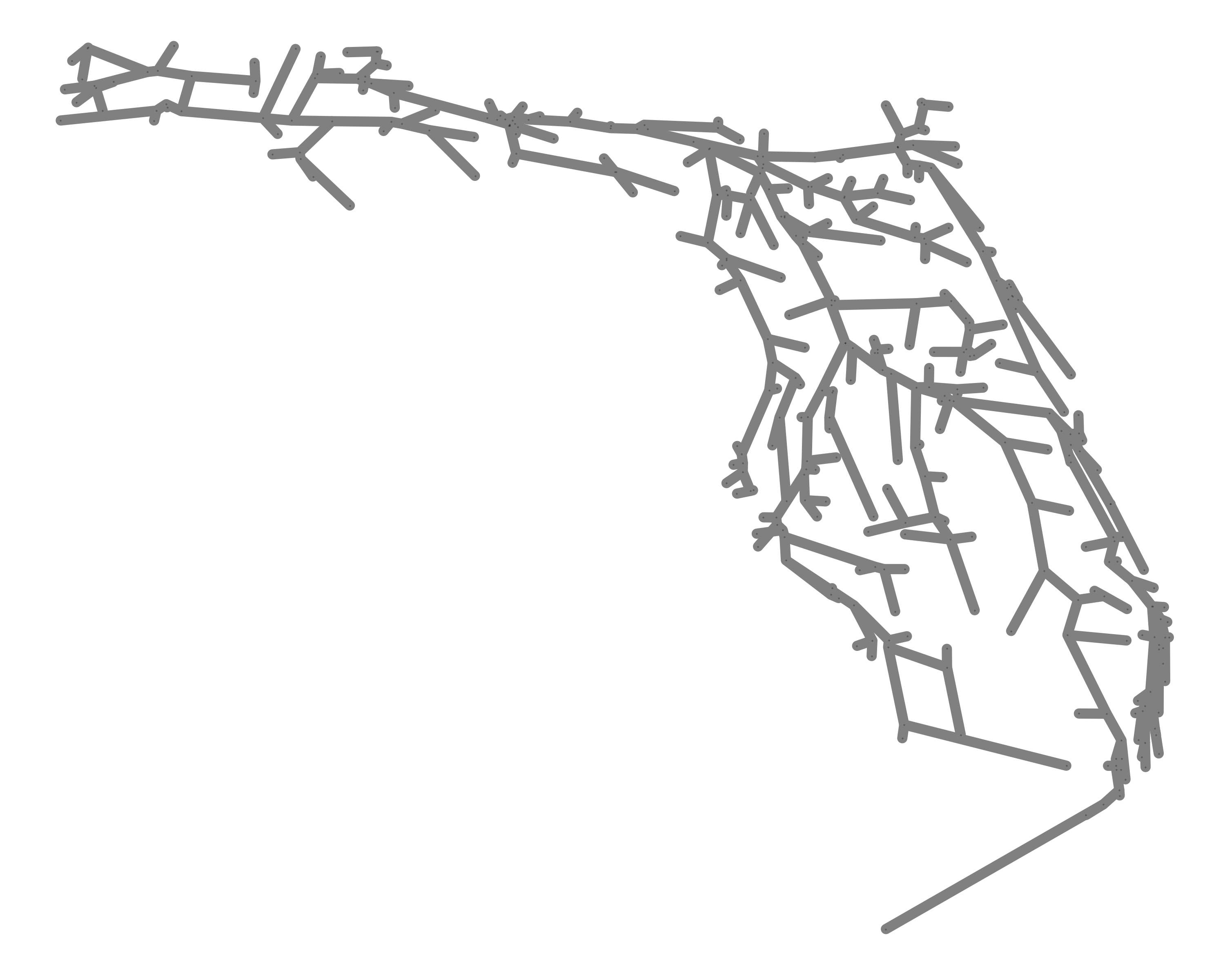}}
        \caption{The traffic network showing the roads across the state of Florida.}
\end{subfigure}\;
\begin{subfigure}[t]{.3\textwidth}
   \centering
   \includegraphics[width=5.4cm, height=3.7cm]{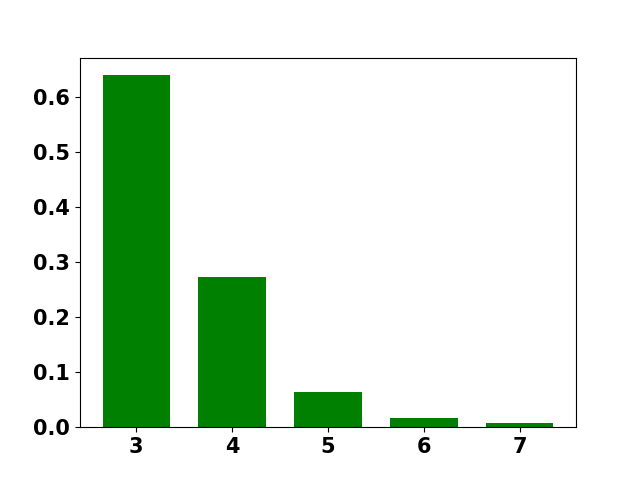}
   \caption{The posterior distribution of the number of non-zero factors.}
\end{subfigure}\;
\begin{subfigure}[t]{.3\textwidth}
    \centering
    \includegraphics[width=4.9cm,height=3.7cm]{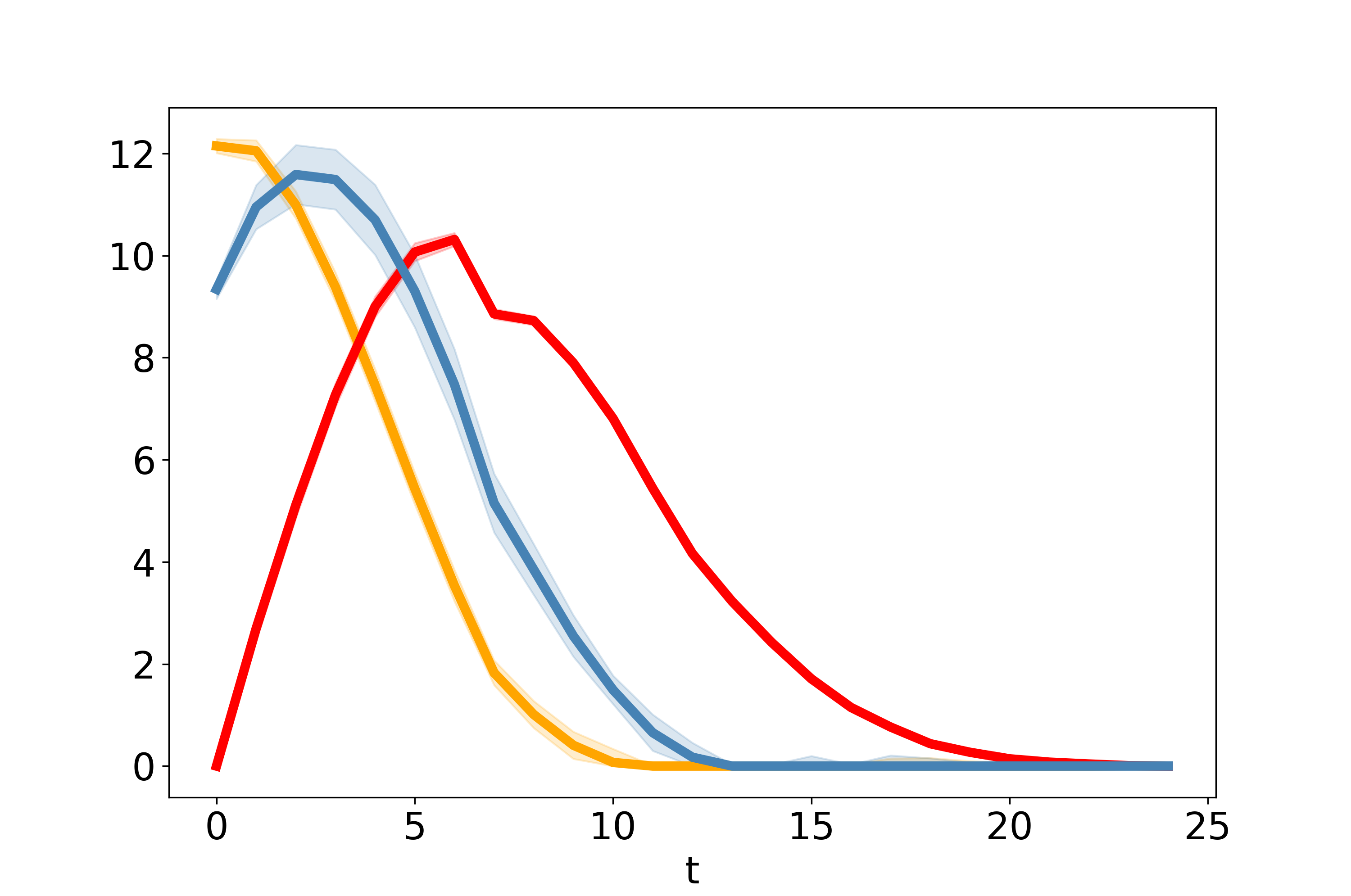}
        \caption{The estimated three loadings $\gamma_l^{(t)}$ changing over time.}
        \vspace{.5cm}
\end{subfigure}
        \begin{subfigure}[t]{.24\textwidth}
\centering
\setlength{\fboxrule}{.1pt} 
\fbox{\includegraphics[width=3.5cm, height=2.5cm]{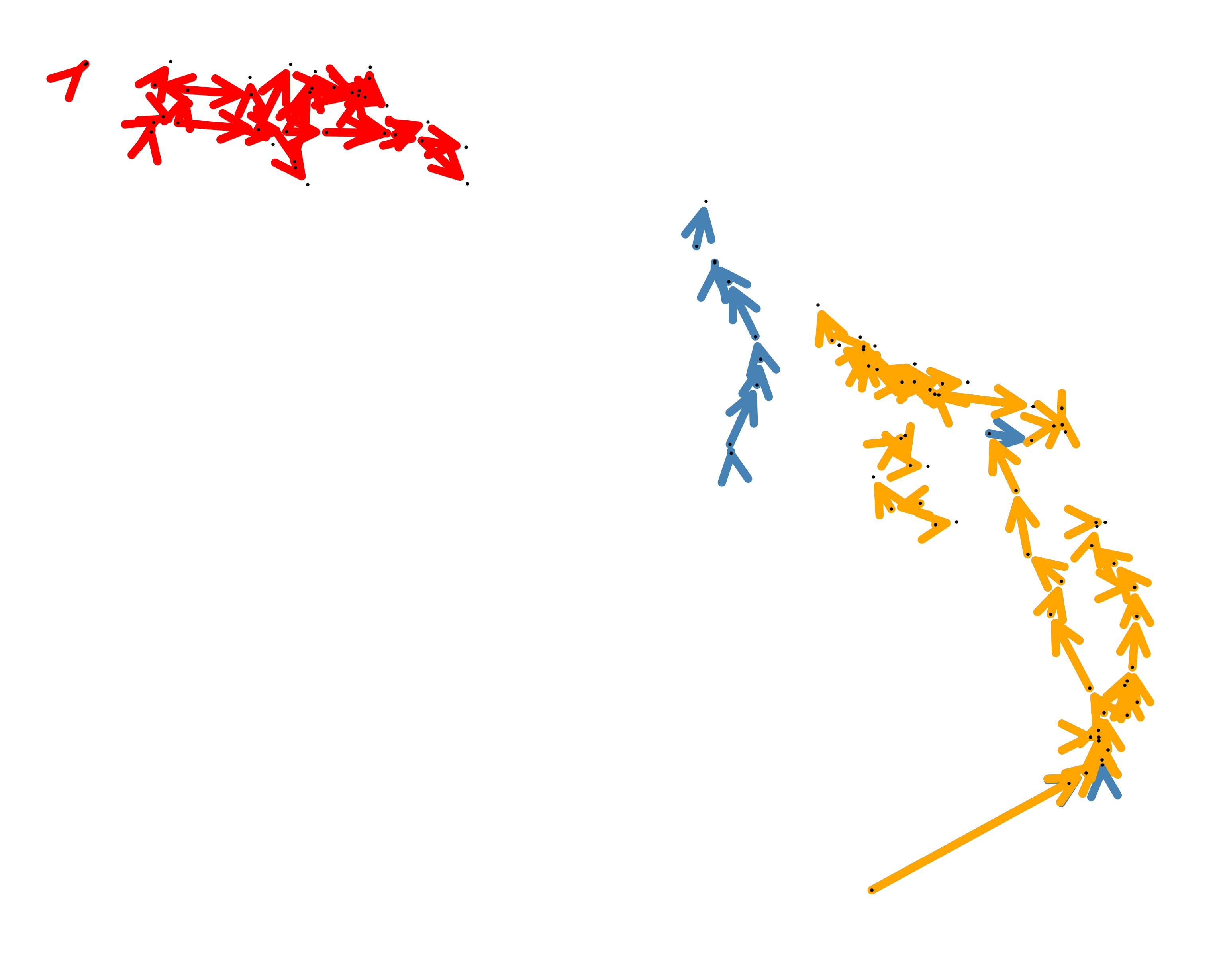}}
        \caption{}
        \end{subfigure} 
\begin{subfigure}[t]{.24\textwidth}
\centering
\setlength{\fboxrule}{.1pt} 
\fbox{\includegraphics[width=3.5cm, height=2.5cm]{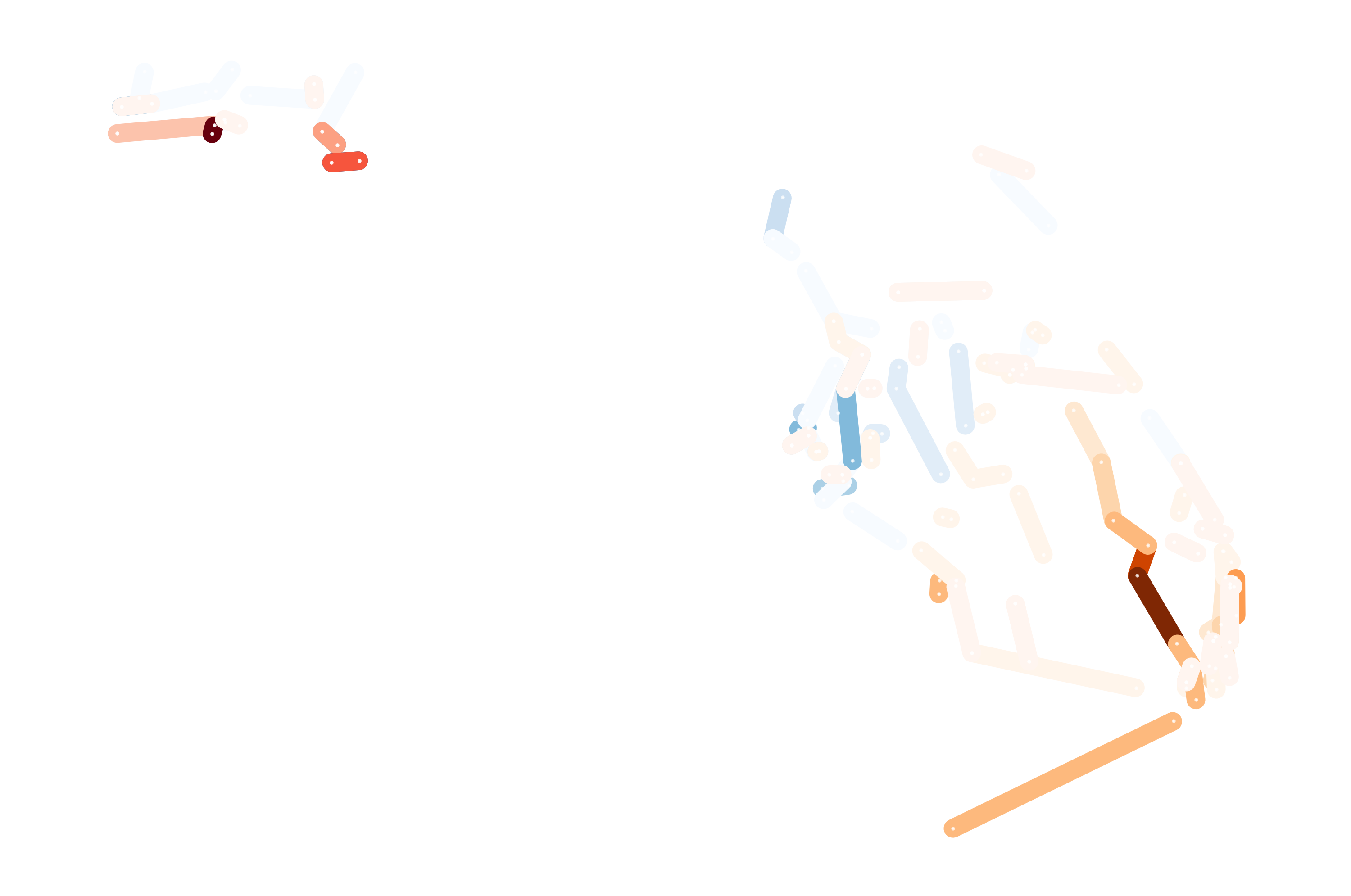}}
        \caption{}
\end{subfigure}
\begin{subfigure}[t]{.24\textwidth}
\centering
\setlength{\fboxrule}{.1pt} 
\fbox{\includegraphics[width=3.5cm, height=2.5cm]{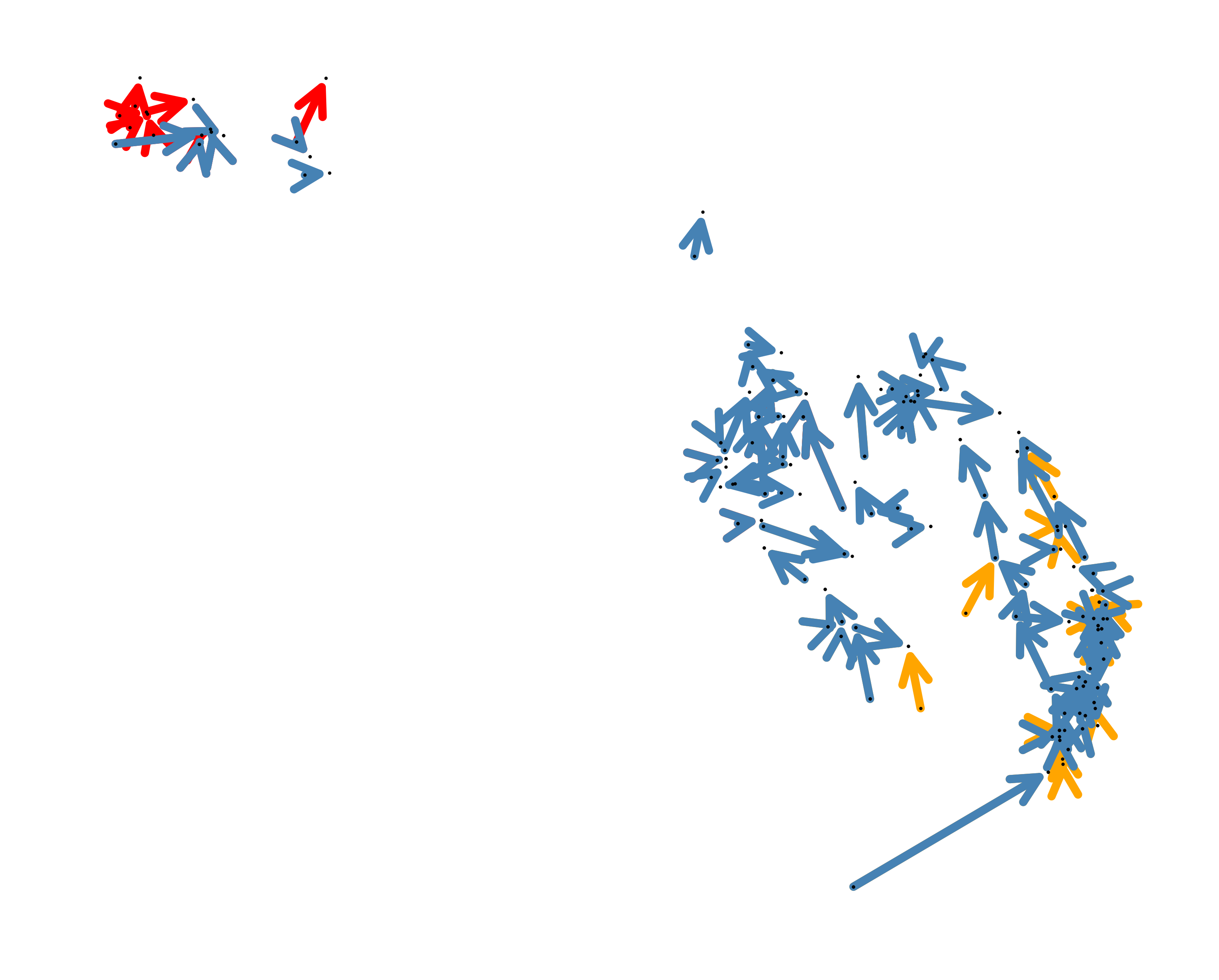}}
       \caption{}
\end{subfigure}
\begin{subfigure}[t]{.24\textwidth}
\centering
\setlength{\fboxrule}{.1pt} 
\fbox{\includegraphics[width=3.5cm, height=2.5cm]{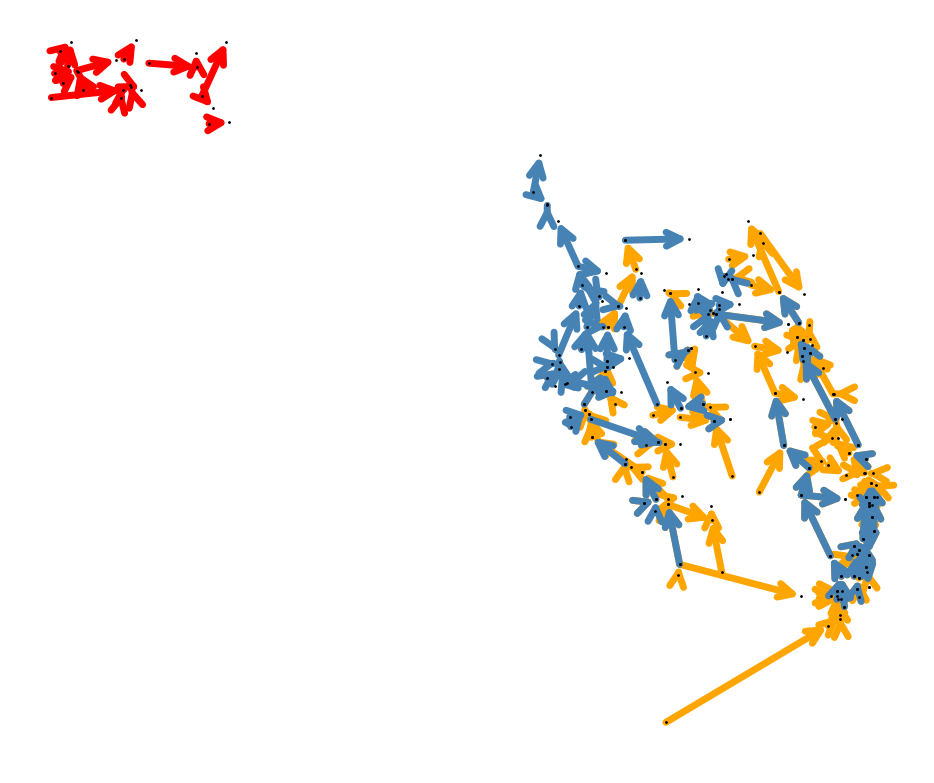}}
       \caption{}
\end{subfigure}
\caption{\label{fig: map} Data analysis on the dynamic flow network, observed during the Hurricane Irma evacuation. A latent factor model is fitted to the data, with the factors regularized by a proximal prior that force each network to be sparse in both the flows and the number of in-flows and out-flows. (d) shows three factor flows (posterior mode) estimated using the proximal prior. Each factor is close to a connected sub-flow network with (e) element-wise variance representing the uncertainty on the factor estimate. We compare the proximal prior with two realizations of shrinkage priors, where (f) shows the factors estimated using elementwise horseshoe prior, which are very fragmented and difficult to interpret. (g) shows the factors estimated using the row-sparse horseshoe, which contain many nodes with small or no flows.}
\end{figure}

Now, to make the factors useful in interpretation, we require each $F^{(l)}$ to be a {\em feasible flow} --- an idealized flow network satisfying the following constraints, 
 (i) skew-symmetry (except for the diagonal): $F^{(l)}_{i,j}=-F^{(l)}_{j,i}$ for $i<j$; (ii) flow-conservation, that the net sum of in-flows should equal to the out-flows for node $j$, $\sum_{i=1}^{n_V} F^{(l)}_{i,j} = 0$; (iii) to reduce noise, we assume that the elements of $F^{(l)}$ are sparse.
Further, as we expect that most of the nodes do not have an external in-low/out-flow, we assume that (iv) most of the nodes having $F^{(l)}_{j,j}=\sum_{i\neq j} F^{(l)}_{i,j}=0$.  To obtain the parameter in such a highly constrained space, we use the proximal mapping, with $\beta^{(l)} \in \mathbb{R}^{n_V\times n_V}$,
   \bel
  \label{eq: prox_network}
  F^{(l)} &=\text{prox}\{ \beta^{(l)} \} = \argmin_{z\in \mathbb{R}^{n_V\times n_V}}    \lambda_1 \|z\|_1 + \lambda_1 \sum_{j=1}^{n_V}|z_{j,j}|+ \frac{1}{2}\|\beta^{(l)}-z\|^2_2 ,\\
  &\text{subject to  } z_{i,j} = -z_{j,i} \text{ for } i\neq j \text{ and }\sum_{i=1}^{n_V}z_{i,j} = 0 \text{ for } j = 1,\ldots,n_V.
  \eel
  The proximal mapping does not have a closed-form solution, however, can be efficiently computed  using the the  alternating direction method of multipliers. 
 Note that with constraint (i) and (ii), it is sufficient to use the lower-triangular entries to represent the rest. In the following, we use $(\beta_{i,j})_{i>j}$ to denote the $n_V(n_V-1)/2$ vector containing the lower-triangular entries. Therefore (17) is equivalent to
   \bel
  \label{eq: prox_network_lower}
  \text{prox}\{(\beta_{i,j})_{i>j}\} = &\argmin_{\{z_{i,j}\}_{i>j}} \frac{1}{2}\sum_{i>j}(\beta_{i,j}-z_{i,j})^2  +  \lambda_1 \sum_{i>j} |z_{i,j}| + \lambda_2 \sum_{j=1}^{n_V}|\sum_{i>j}z_{i,j}-\sum_{i<j}z_{j,i}|.
  \eel  
 This proximal operator is evaluated via the  alternating direction method of multipliers and solved iteratively. We provide the detailed algorithm in the appendix.

  On the prior of the loading, we assign a group shrinkage prior by using the $(2,1)$-matrix norm in the proximal mapping. For the matrix $\gamma\in \mathbb{R}^{d\times T}$, we set:
 \[
 \gamma= \text{prox}_{\lambda_2\|\cdot\|_{2,1}}(\rho) = \argmin_{ z\in\mathbb{R}^{d\times T}} 
 \lambda_2 \|z\|_{2,1}+ \frac{1}{2}
\|\rho -z\|_F^2 ,
 \]
 where  $\|z\|_{2,1}=\sum_{l=1}^{d}\sqrt{\sum_{t=1}^{T}\{ z_{l}^{(t)}\}^2}$. This prior has the advantage that $\{\gamma_l^{(1)}, \ldots, \gamma_l^{(T)}\}$ will be simultaneously  zero for certain $l$ --- which allows us to use an overfitted model with a relatively large $d=10$, with the posterior recovering only a small number of factors with non-zero loadings. We use independent standard normal as prior on the elements of $\beta^{(l)}$ and $\rho$.
 
 We run the HMC for 20,000 steps and discard the first 5,000 as burn-ins, and we use thinning at every 10th iteration as the posterior sample. The posterior shows the highest probability at having $3$ factors, and we visualize them in  Figure \ref{fig: map}(d) --- clearly, by forcing the external in-flows and out-flows to be sparse, we have each factor roughly corresponding to a single connected sub-network.

 Interestingly, examining the estimated loadings that change over the time points in Figure \ref{fig: map}(c), we see that in the beginning of the evacuation, the factors 1 and 2 are dominant, but later there is a sudden decrease --- this in fact corresponds to the time point when the hurricane makes the landfall,  effectively forcing the traffic in those areas to shut down. After the 60th hour, the traffic moves up to the north part, and factor 3 represents the late stage of the evacuation.

 To compare, we also test two continuous shrinkage priors on the factors. Specifically, we use  (a) the elementwise horseshoe prior on each lower-diagonal $F^{(l)}_{i,j} \sim \text{N}(0, \tau^2_{i,j}\sigma^2)$, $\tau_{i,j} \sim C^+(0,1)$, $\sigma^2\sim \t{Inverse-Gamma}(2,0.01)$, and (b) the two-way group horseshoe prior, by letting $F^{(l)}_{i,j}\sim \text{N}(0, \tau_i\tau_j\sigma^2)$,
$\tau_{i} \sim C^+(0,1)$, $\sigma^2\sim \t{Inverse-Gamma}(2,0.01)$. The purpose of (b) is to shrink each row of $F$ simultaneously, while satisfying the skew-symmetry of $F^{(l)}$. Effectively these horseshoe priors accommodate the properties of (i)(ii)(iii) of a sparse feasible flow, nevertheless, they cannot accommodate (iv)
--- as each $F^{(l)}_{j,j}$ is completely determined given $F^{(l)}_{i,j} \; (i\neq j)$, we could not further assign shrinkage prior on $F^{(l)}_{j,j}=\sum_{i\neq j}F^{(l)}_{j,j}$. As the result in Figure \ref{fig: map}(f), the elementwise continuous shrinkage priors show a large number of external in-flows and out-flows, leading to fragmented small networks in each factor. As shown in Figure \ref{fig: map}(g), the two-way group horseshoe finds many nodes with no flows at all, which is not very interpretable since we would like most of the nodes to have many in-flows and out-flows, as long as the total net-flow is zero.

 \vspace{-.5cm}
\section{Discussion}
\label{sec: discuss}
In this article, we exploit the proximal mapping to produce a new class of priors. As we have demonstrated, these priors and the associated probabilistic models can enable statistical inference (such as uncertainty quantification, hypothesis testing) on a wide range of problems, where in the past, one has been limited to point estimate only. The technique of ``data augmentation using optimization'' we have introduced could be generalized for other purposes, such as potentially new efficient algorithm for the posterior computation. Lastly, one could consider other type of optimization problems for a similar prior construction, such as the popular classes of semi-definite \citep{vandenberghe1996semidefinite} and / or mixed integer programmings \citep{linderoth1999computational}, although how to provide a probabilistic treatment for these problems is still an open question.

\appendix

\section{Appendix}

\subsection{Proof of Theorem 1}
\begin{proof}
For any value $\beta^*:\t{prox}_{\lambda g_\gamma}(\beta^*)=\theta^*$ using $g_\gamma $ at given $\gamma$ and $\lambda$, $\epsilon<\|\t{prox}_{\lambda g_\gamma}(\beta)-\t{prox}_{\lambda g_\gamma}(\beta^*)\|\le \|\beta-\beta^*\|$. Therefore,  ${\bf 1}(\theta, \|\theta-\theta^*\|>\epsilon\mid y, \lambda, \gamma) \le {\bf 1}\left(\beta,\left\|\beta-\beta^{*}\right\|>\epsilon \mid y  ,\lambda, \gamma\right).$ Taking the minimum over $\beta^*$ on the right hand side and expectation on both sides, we obtain  the first result.

Next, using the fact that for two independent copies $\beta_1$, $\beta_2$ from $\Pi(\beta\mid y)$, $2\t{tr}[Cov(\beta\mid y)= \mathbb{E}_{\bm \beta_1,\bm \beta_2}(\|\beta_1-\beta_2\|_2^2 \mid y)
=\mathbb{E}_{\bm \lambda,\bm \gamma} \mathbb{E}_{\bm \beta_1,\bm \beta_2} (\|\beta_1-\beta_2\|_2^2 \mid \lambda, \gamma,y),
$
and the non-expansiveness of proximal mappings, we obtain the second result.
\end{proof}

\subsection{Proof of Theorem 2}
\begin{proof}
Let  $0 < \lambda_{1}<\lambda_{2}$, ${v}_{1}=\operatorname{prox}_{\lambda_{1} g}({x})$and ${v}_{2}=\operatorname{prox}_{\lambda_{2} g}({x}),$ we prove: (i) $g(v_1)\ge g(v_2)$ and (ii) $\|v_1-x\|\le \|v_2-x\|.$
 For (i),
$$
\begin{aligned}
 \frac{1}{2}\left\|{v}_{2}-{x}\right\|^{2}+\lambda_{2}g\left({v}_{2}\right) 
=& \frac{1}{2}\left\|{v}_{2}-{x}\right\|^{2}+\lambda_{1}g\left({v}_{2}\right)+\left(\lambda_{2}-\lambda_{1}\right)g\left({v}_{2}\right) \\
\stackrel{(a)}\geq & \frac{1}{2}\left\|{v}_{1}-{x}\right\|^{2}+\lambda_{1}g\left({v}_{1}\right)+\left(\lambda_{2}-\lambda_{1}\right)g\left({v}_{2}\right) \\
=& \frac{1}{2}\left\|{v}_{1}-{x}\right\|^{2}+\lambda_{2}g\left({v}_{1}\right)+\left(\lambda_{2}-\lambda_{1}\right)\left \{ g\left({v}_{2}\right)-g\left({v}_{1}\right)\right\} \\
\stackrel{(b)}\geq & \frac{1}{2}\left\|{v}_{2}-{x}\right\|^{2}+\lambda_{2}g\left({v}_{2}\right)+\left(\lambda_{2}-\lambda_{1}\right)\left \{ g\left({v}_{2}\right)-g\left({v}_{1}\right)\right\},
\end{aligned}
$$
where $(a)$ is due to $v_1$ is the minimizer of $\frac{1}{2}\left\|{v}-{x}\right\|^{2}+\lambda_{1}g\left({v}\right)$,  and similarly for $v_2$ in $(b)$.
Therefore, $\left(\lambda_{2}-\lambda_{1}\right)\left \{ g\left({v}_{2}\right)-g\left({v}_{1}\right)\right\} \leq 0$, which leads to $g\left({v}_{1}\right) \ge g\left({v}_{2}\right).$ 

For (ii), slightly changing (a), we have
$
% \frac{1}{2}\left\|{v}_{2}-{x}\right\|^{2}-\lambda_{2}\left(g\left({v}_{1}\right)-g\left({v}_{2}\right)\right)\le
\frac{1}{2}\left\|{v}_{1}-{x}\right\|^{2} \le\frac{1}{2}\left\|{v}_{2}-{x}\right\|^{2}-\lambda_{1}\left\{g\left({v}_{1}\right)-g\left({v}_{2}\right)\right \}.
$
Since $\lambda_1\ge 0$, we have $\left\|{v}_{1}-{x}\right\|^{2}\le\left\|{v}_{2}-{x}\right\|^{2} $.
\end{proof}

\subsection{Proof of Theorem 3}
\begin{proof}
         Since the proximal mapping satisfies $1$-Lipschitz condition,  $\t{diam}\{\t{prox}(B)\}\le \t{diam} (B)$ for all $B$ in the domain of $\prox$. Using the definition of the Hausdorff measure, $\mathcal H^s\{\t{prox}(A)\}\le \mathcal H^s(A)$. To see the statement 2, we apply the result in statement 1 and see $ \mathcal H^s\{\t{prox}(\mathcal A)\}=0$ whenever $\mathcal  H^s(\mathcal A)=0$.
\end{proof}

\subsection{Proof of Theorem 5}
\begin{proof}
Using Theorem 3.2.22 of \cite{federer2014geometric}, for any $\mathcal H^{p}$ measurable function $F$ on $\Beta^k$,
\be
\int_{\Beta^k} F(\beta) J_{k} \prox(\beta) \mathcal H^p( \textup d \beta)
= \int_{\Theta^k} \int_{\prox^{-1}(\theta)}F(b)  \mathcal H^{p-k}( \textup d b) \mathcal H^{k}( \textup d \theta).
\ee
Using the assumption, we can exclude the zero-measure set where  $J_{m_k} \prox(\beta)=0$.

\end{proof}

\subsection{Algorithm to Compute the Proximal Mapping in the Flow Network Modeling}

  We formulate an equivalent problem to \eqref{eq: prox_network_lower} 
   \bel
  \label{eq: eqprox_network} 
   \text{prox}\{(\beta_{i,j})_{i>j}\} = &\argmin_{\{z_{i,j}\}_{i>j},x\in\mathbb{R}^{n_V}} \frac{1}{2}\sum_{i>j}(\beta_{i,j}-z_{i,j})^2  +  \lambda_1 \sum_{i>j} |z_{i,j}| + \lambda_2 \sum_{j=1}^{n_V}|x_j|\\
    &\text{subject to  } C(z_{i,j})_{i>j} = x,
  \eel  
  where $C\in\mathbb R^{n_V\times n_V(n_V-1)/2}$ is the matrix such that $\{C(z_{i,j})_{i>j}\}_k = \sum_{i>k}z_{i,k}-\sum_{i<k}z_{k,i}$.
  The scaled augmented Lagrangian for \eqref{eq: eqprox_network} is: 
  \[
  \mathcal L = \frac{1}{2}\|(\beta_{i,j})_{i>j}-(z_{i,j})_{i>j}\|^2_2 + \lambda_1\|(\beta_{i,j})_{i>j}\|_1+\lambda_2\|x\|_1 + \frac{\gamma}{2}\|C(z_{i,j})_{i>j}-x+u\|_2^2 - \frac{\gamma}{2}\|u\|_2^2.\]
  
  In each iteration we update the $x$ and $z$ separately to minimize the Lagrangian:
  \be
 ( z_{i,j})_{i>j}&\leftarrow \text{prox}_{\lambda_1 / \gamma\|\cdot\|_1}[(I+\gamma C^{\textrm{T}}C)^{-} \{\gamma C^{\textrm{T}} (x-u)+\beta\}];\\
  x&\leftarrow \text{prox}_{\lambda_2 / \gamma\|\cdot\|_1}\{C(z_{i,j})_{i>j}+u\},
  \ee 
  and update $u$ as in dual ascent:
  \be
  u \leftarrow u + C(z_{i,j})_{i>j} - x
  \ee 
  until convergence (i.e., $\|C(z_{i,j})_{i>j}-x\|\to 0$).

% \subsection{\maoran{Convergence of HMC}}
% To show that the HMC converges fast to the high posterior region and does not depend on the initialization, we run 10 chains with different initializations in a sparse model. We consider a high-dimensional regression problem $$y_i = x_i\theta+\epsilon_i,i=1,\ldots,n,$$  where $x_i\stackrel\text{iid}{\sim} N(0,I_{p})$ with $n = 200, p = 500$, $\theta$ has the first five entries being $5$, and other entries being $0$, the error term $\epsilon\sim N(0,I_{p})$. 

% Figure \ref{figure: convergeMCMC} shows ten trajectories of log-posterior probability and $theta_1$ versus the
% number of iterations of the Markov chain. The chains converge quickly to stationary, with high probability regions near the true model.
% \begin{figure}[htbp]
%     \centering
% \begin{subfigure}[t]{.45\textwidth}
% \centering
% \includegraphics[width=\textwidth]{figure/mcmc_200500_pos.png}
% \caption{Trace plots of the log-posterior probability.}
% \end{subfigure}\qquad
% \begin{subfigure}[t]{.45\textwidth}
% \centering
% \includegraphics[width=\textwidth]{figure/mcmc_200500_theta1.png}
% \caption{Trace plots of the first entry of $\theta$, where the truth is $5$.}
% \end{subfigure}
% \caption{\label{figure: convergeMCMC}Mixing performance of HMC with ten perturbations of the true model. Each curve corresponds to one trajectory of the chain.}
% \end{figure}

\spacingset{1.2} % DON'T change the spacing! 

\bibliographystyle{natbib}

\end{document}